\documentclass[journal]{IEEEtran}
\usepackage[utf8]{inputenc}
\usepackage{tabularx}
\usepackage{cite}
\usepackage{amsmath,amssymb,amsfonts}
\usepackage{graphicx}
\usepackage[flushleft]{threeparttable}
\usepackage{comment}
\usepackage{graphicx}
\usepackage{subcaption}
\usepackage{amsmath}

\usepackage[
  left=0.54in,
  right=0.54    in,
  top=0.65in,
  bottom=1in
]{geometry}

\usepackage{colortbl}
\usepackage{xcolor}

\usepackage{textcomp}
\usepackage[table]{xcolor}
\usepackage{subcaption}
\usepackage{gensymb}
\usepackage{float}
\usepackage{algorithm,algcompatible,algpseudocode}
\usepackage{array}
\usepackage{hyperref}
\usepackage{eso-pic, rotating, graphicx}
\usepackage{multirow}
\usepackage{multicol}
\usepackage{tabularx, makecell}
\usepackage[utf8]{inputenc}
\usepackage{tabularray}
\usepackage{booktabs}
\usepackage{algpseudocode}
\usepackage{glossaries}

\newacronym{cl}{CL}{Contrastive Learning}
\newacronym{dc}{DC}{Deep Clustering}
\newacronym{ae}{AE}{Auto Encoder}
\newacronym{rf}{RF}{Radio Frequency}
\newacronym{dmm}{DMM}{Decision Making Module}
\newacronym{sei}{SEI}{Specific Emitter Identification}
\newacronym{ued}{UED}{Unknown Emitter Detection}
\newacronym{fe}{FE}{Features Extractor}
\newacronym{oran}{O-RAN}{Open Radio Access Network}
\newacronym{dm}{DMS}{Different Messages Scenario}
\newacronym{kan}{KAN}{Kolmogorov-Arnold Network}
\newacronym{flm}{FLM}{Feature Learning Module}
\newacronym{ml}{ML}{Machine Learning}
\newacronym{sm}{SMS}{Same Messages Scenario}
\newacronym{ssl}{SSL}{Self-Supervised Learning}
\begin{document}
\pagestyle{plain}

\title{Design Principles of Zero-Shot Self-Supervised Unknown Emitter Detectors}


\author{\IEEEauthorblockN{Mikhail Krasnov\IEEEauthorrefmark{1}\IEEEauthorrefmark{2}, Ljupcho Milosheski\IEEEauthorrefmark{1}\IEEEauthorrefmark{2},
Mihael Mohorčič\IEEEauthorrefmark{1}\IEEEauthorrefmark{2}, Carolina Fortuna\IEEEauthorrefmark{1}}\\
\IEEEauthorblockA{\IEEEauthorrefmark{1} Jožef Stefan Institute, Ljubljana, Slovenia\\
\IEEEauthorrefmark{2}Jožef Stefan International Postgraduate School, Ljubljana, Slovenia\\
Email: \{mikhail.krasnov, ljupcho.milosheski, miha.mohorcic, carolina.fortuna\}@ijs.si }}

\maketitle
\begin{abstract}

Robust situational awareness in contested wireless environments requires the ability to detect unauthorized or hostile emitters without prior knowledge of their signatures. Existing studies on \acrlong{ued} (\acrshort{ued}) and identification are hindered by their reliance on labeled or proprietary datasets, unrealistic assumptions (e.g., all samples containing identical transmitted messages), or a lack of systematic evaluation across architectures and design dimensions.
In this work, we identify the \textit{design dimensions} of machine learning based \acrshort{ued} and introduce a \textit{structured workflow} for developing such systems. The workflow, suitable for guiding studies and end-to-end label-free \acrshort{ued} designs, consists of four key components: (a) data modality, (b) feature learning module, (c) machine learning approach, and (d) decision-making module. Using this workflow, we examine the design principles of \acrshort{ued} in two distinct  transmission scenarios: same messages scenario (SMS) and different messages scenario (DMS). 
Our investigation yields several key findings. (a) Under realistic  DMS, the 2D constellation data modality achieves up to a 20-percentage-point improvement in ROC-AUC compared to the conventional raw I/Q representation. (b) Kolmogorov-Arnold Networks (KANs) provide interpretable representations while achieving performance comparable to CNNs in both scenarios (c) for DMS, the best-performing configuration combines SVD-initialized KANs with a deep clustering approach and the 2D constellation modality, improving ROC-AUC by up to 20 percentage points over standard KANs with identical workflow components. (d) through analysis of the decision-making module, we determine the optimal number of clusters for environments with varying numbers of known emitters. 
\end{abstract}

\section{Introduction}
\label{sec:intro}

\begin{figure}[!t]
\centering
\hfill
\centering
\begin{subfigure}[b]{0.5\textwidth}
    \centering
    \includegraphics[width=0.7\columnwidth]{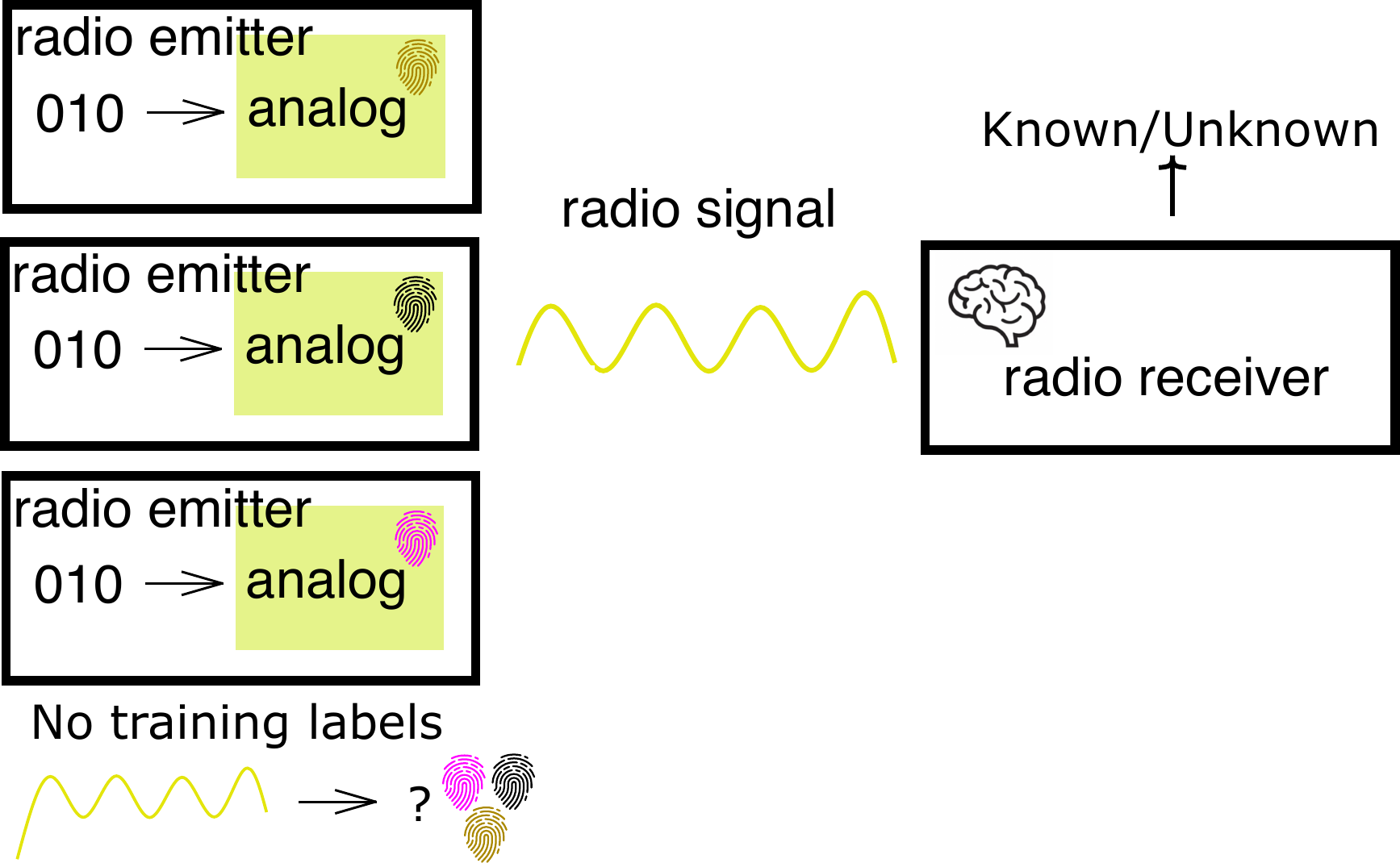}
    \caption{The concept of self-supervised detection and identification of emitters.}
    \label{fig:problem}
\end{subfigure}
\centering
\begin{subfigure}[b]{0.48 \textwidth}
    \includegraphics[width=\columnwidth]{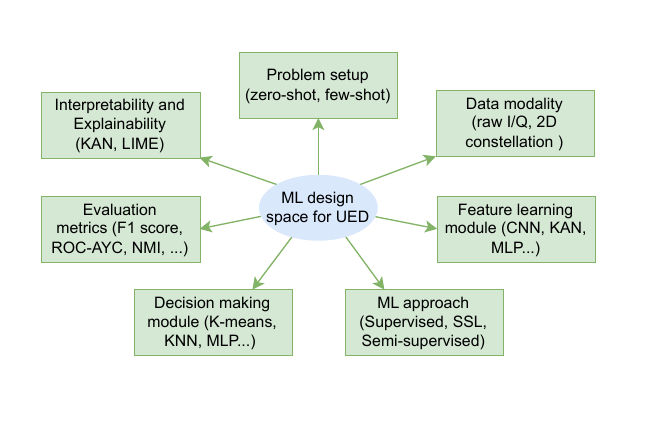}
    \caption{Design space for building deep learning model for detection and identification of emitters.}
    \label{fig:design_choices}
\end{subfigure}
\hfill

\caption{The design space for \acrshort{ml}  detection and identification of known/unknown emitters.}
\label{fig:2D_const}
\end{figure}

The management and operation of wireless communication networks increasingly depend on situational and spectrum awareness to dynamically adapt to environmental variations and user/application-imposed demands. Spectrum awareness and network security are envisioned as part of the main research directions \cite{polese2023understanding} for the future Open Radio Access Networks (O-RAN) \cite{alliance2020ran}, but the same techniques can also be used in interference management, regulatory compliance and spectrum optimization. Together with the development of modulation and technology classification \cite{milosheski2023self}, they contribute to the creation of efficient, spectrum-aware, wireless communication networks \cite{ouyang2023integrated}. One of the key challenges related to spectrum awareness lies in the reliable detection and identification of emitters within the communication range, on the conceptual level depicted in Figure~\ref{fig:problem}. It relies on robust signal processing techniques and increasingly leverages advanced machine learning approaches. In this respect, the research community distinguishes between two complementary processes within the broader context of devices recognition: the Specific Emitter Identification (\acrshort{sei}) and the Unknown Emitter Detection (\acrshort{ued}). 

\acrshort{sei} focuses on recognizing and classifying known emitters under a closed-set assumption based on unique, device-specific signal alterations such as transient behaviors, hardware-induced imperfections, or modulation patterns allowing precise attribution of transmissions to individual sources \cite{ding2018specific}. This makes \acrshort{sei} particularly well suited for network security applications \cite{robinson2020dilated}. The aim is to distinguish emitters that are allowed to operate in a certain wireless network, in which case they are transmitting useful information and can be referred to as transmitters. In contrast, \acrshort{ued} \cite{apfeld2021recognition} is working under open-set assumption and it aims to identify signals that deviate from any known emitter profile, detecting and isolating novel, spoofed, or adversarial sources that fall outside the established feature space of the known emitters. In network security application, \acrshort{ued} is used to detect and reject or flag unrecognized signals to prevent their misclassification by \acrshort{sei}. As such, these processes form a closed-loop detection and identification system whereby  \acrshort{sei} provides fine-grained, instance-level identification within a known feature space used for training, whereas \acrshort{ued} ensures robustness and adaptability by flagging outliers and prompting model updates or human review for inclusion of novel devices, together maintaining situational awareness and security in dynamic or contested radio environments.

Most of the existing approaches are focusing on the closed-set classification problem. Such approaches consider that all the emitters are known and labeled transmission data is available for each of them for developing supervised classification models \cite{wong2018clustering, han2022proceed}, reflecting the authentication use case of known devices. Some of these models achieved remarkable performance in distinguishing known devices \cite{luo2022transformer,basak2021combined,wang2021specific}, even when thousands of such devices \cite{robinson2020dilated} exist in the network. However, in real-world operating scenarios with the rising volume and types of wireless devices, new transmitters could appear regularly in the network, which could pose significant security and management challenges. In such open-set setup there is little or no labeled data available for training, making the supervised learning for model development unsuitable. The open-set challenges are divided into two distinct subsets: few-shot learning \cite{xu2023few, huang2022deep, liu2023overcoming} and zero-shot learning \cite{robinson2021novel}. Few-shot learning follows the supervised evaluation protocol but operates under severe label constraints, with the number of labeled samples ranging from just a few (1–20) \cite{huang2022deep} to several hundred \cite{peng2023supervised}. In contrast, zero-shot learning aims to detect emitters that are entirely absent during model training \cite{dong2021sr2cnn,wen2022drsn}. 

We notice that, overall, the approach to developing a system  capable of learning to distinguish emitters based on device-specific signal alterations largely depends on the application and systematic consideration of all the existing design choices, as visualized in Figure \ref{fig:design_choices}. The problem set-up considering zero-shot and few shot learning is one dimension. Other dimensions relate to data modality, feature learning, \acrshort{ml} approach, etc. The selection of data modality may significantly improve the final performance, but it also influences the complexity of the feature learning module. The design of the feature learning module influences many of the other design choices, such as performance, complexity, explainability, and interpretability. The existing model architectures such as Convolutional Neural Networks (CNN) \cite{robinson2021novel}, transformers \cite{luo2022transformer}, Long Short-Term Memory (LSTM) recurrent neural networks \cite{al2021deeplora} and/or combinations with Multilayer Perceptron (MLP) \cite{stankowicz2021unsupervised} have been already proven in related domains and they usually provide satisfactory performance \cite{robinson2021riftnet}. However, designing a dedicated model for the radio signal specifics and predefined use cases can lead to significant complexity reduction \cite{milosheski2024spectrum}. 

The choice of the neural architecture for feature learning also affects its explainability and interpretability. For the deep learning models, there are several explainability approaches, such as Local Interpretable Model-agnostic Explanations (LIME) \cite{ribeiro2016should}, SHapley Additive exPlanations) (SHAP) \cite{lundberg2017unified}, and saliency maps \cite{simonyan2013deep}, whereas the issue of model interpretability remains largely unaddressed. This is primarily due to the complexity and opacity of the deep learning modules employed, typically including large CNNs or self-attention-based architectures \cite{luo2022transformer}, which make it difficult to understand or explain model decisions, while also lacking interpretability. So far, the interpretable models are the classic machine learning (ML) approaches, such as Support Vector Machine (SVM), and logistic regression \cite{riyaz2018deep}, which are significantly outperformed by the deep learning models, while interpretability is overlooked in most of the recent works. However, due to the potential use in security-related applications, such as device authentication, interpretability of the detection and identification of emitters remains an important design choice. 

For zero-shot set-ups,  \acrfull{ssl} \cite{liu2021self} techniques are the most suitable ML approaches. \acrshort{ssl} represents a family of machine learning methods, which create supervision signal without using ground truth labels but through data manipulation techniques such pseudo-labeling, augmentations, reconstruction etc. 
By formulating and solving pretext tasks derived directly from the data, \acrshort{ssl} enables the development of models that generalize well without relying on manual labeling. This makes it particularly well-suited for open-set detection and identification of emitters \cite{peng2023supervised, stankowicz2021unsupervised, zhang2023transmitter}, closely aligning with realistic deployment conditions, where the diversity of emitters and the dynamic nature of the spectrum make comprehensive labeling impractical.

\begin{table*}[t!]
\centering
\begin{threeparttable}
\caption{Comparison of related works on \acrshort{sei} and \acrshort{ued} tasks with focus on semi-supervised, unsupervised and self-supervised learning approaches}
\label{tab:combined_sei_ued_summary}
\renewcommand{\arraystretch}{1.2}
\begin{tabular}{|p{0.6cm}|p{3.4cm}|p{1.9cm}|p{2.3cm}|p{1.2cm}|p{1.6cm}|p{2.7cm}|p{1.3cm}|}
\hline
\textbf{Ref.} & \textbf{Problem Setup} & \textbf{Data Modality} & \textbf{\acrshort{ml} Approach} & \textbf{Feature Extractor} & \textbf{Interpret / Explain} & \textbf{Supervision Setup} & \textbf{Results / Notes} \\ \hline \hline

\cite{hao2023contrastive} & Self-supervised \acrshort{sei}/\acrshort{ued} & Raw I/Q & \acrshort{cl} & CNN & \acrshort{dmm} level & No. of training emitters & 93\% F1 \\ \hline

\cite{liu2023overcoming} & Self-supervised \acrshort{sei} & Raw I/Q & \acrshort{cl} (Viewmaker) & CNN & None & Few-shot fine-tuning & 83\% ACC \\ \hline

\cite{krasnov2025novel} & Unsupervised \acrshort{sei} & Raw I/Q & \acrshort{dc} & CNN & \acrshort{dmm} level & Unsupervised & 83\% F1 \\ \hline

\cite{huang2022deep} & Masked \acrshort{ae} (Reconstruction) & Raw I/Q & \acrshort{ae} & CNN &  None & Few-shot fine-tuning & 83\% ACC \\ \hline

\cite{xu2023few} & Hybrid Few-shot \acrshort{sei} & Raw I/Q & Hybrid (\acrshort{ae} + \acrshort{cl}) & CNN & None & Few-shot fine-tuning & 90\% ACC \\ \hline

\cite{yao2023few} & Few-Shot \acrshort{sei} & Raw I/Q & Assymetric M\acrshort{ae} & CNN & None & Few-shot fine-tuning & 93\% ACC \\ \hline

\cite{wu2023specific} & Few-Shot \acrshort{sei} & Raw I/Q & SimCLR & CNN & None & Few-shot fine-tuning & 70\% ACC  \\ \hline

\cite{zha2023cross} & Few-Shot \acrshort{sei} & Raw I/Q & SimSiam \acrshort{cl} & CNN &None & Few-shot fine-tuning & 80\% ACC \\ \hline

\cite{sun2025few} & Few-Shot \acrshort{sei} & Raw I/Q  & \acrshort{cl}  & CNN & None &Few-shot fine-tuning & 95\% ACC \\ \hline

\cite{liu2022specific} & Few-Shot \acrshort{sei} & 2D constellation & SimCLR & CNN &None & Few-shot fine-tuning & 90\% ACC 
\\ \hline
\textbf{Ours} & Self-supervised zero-shot \acrshort{ued} & Raw I/Q, 2D constellation & \acrshort{cl}, \acrshort{dc}, \acrshort{ae} & CNN, \acrshort{kan}& \acrshort{dmm} and \acrshort{fe} levels & Unsupervised & 85\% F1, 94\% ACC \\ \hline

\end{tabular}

\begin{tablenotes}
\item \textbf{\acrshort{cl}} – Contrastive Learning, 
\textbf{\acrshort{dc}} – Deep Clustering, 
\textbf{\acrshort{ae}} – Autoencoder, 
\textbf{\acrshort{dmm}} – Decision-Making Module, 
\textbf{\acrshort{fe}} – Feature Extractor
\end{tablenotes}
\end{threeparttable}
\end{table*}

In this work, we identify the \textit{design dimensions} of machine learning based \acrshort{ued} as depicted in Figure \ref{fig:design_choices}:  1) problem set-up, 2)  data modality, 3) feature learning, 4) ML approach, 5) decision making, 6) evaluation metrics and 7) explainability and interpretability. Focusing on realistic environments where no prior information is available, we consider a \textit{zero-shot} set-up that inherently requires a \acrshort{ssl} approach. Consequently, we introduce a \textit{structured workflow} for developing such systems. The workflow, suitable for guiding studies for end-to-end label-free \acrshort{ued} designs. Using this workflow, we examine the design principles of \acrshort{ued} in two distinct  transmission scenarios: same messages scenario (SMS) and different messages scenario (DMS) enabled by two distinct datasets \cite{sankhe2019oracle, hanna_wisig_2022}. Throughout the study we consider two possible data modalities, three different feature learning and three different \acrshort{ssl} approaches evaluated through several metrics including also explainability and interpretability considerations. For replicability purposes, the scripts are available as open source\footnote{\url{https://github.com/sensorlab/ZeroUED}}. We summarize our contributions as follows:

\begin{itemize}
    \item We formalize a structured, self-supervised workflow for zero-shot UED on unlabeled data, consisting of four key components: a) data modality, b) feature learning module, c) ML approach, and d) decision-making module. 
    
    \item  Through the lens of the proposed workflow, we investigate the design principles of designing a an \acrshort{ued} on two transmission scenarios: \acrfull{sm} and  \acrfull{dm}, two data modalities, two neural architectures, including KAN designed for interpretability, and three \acrshort{ssl} ML approaches.

    \item During our investigation, we find the following. a) Under the realistic \acrshort{dm} scenario, 2D constellation data modality shows up to 20 p.p. ROC-AUC metric improvement compared to conventional raw I/Q representation. b) Kolmogorov-Arnold networks (\acrshort{kan}s) are interpretable and have performance comparable to CNNs, a convenient alternative that suffers from "black-box" design, across \acrshort{sm} and \acrshort{dm}. c) We observe that for \acrshort{dm}, the best performing configuration is SVD initialized \acrshort{kan} with deep clustering approach and 2D constellation data modality. This configuration improves ROC-AUC performance up to 20 p.p. compared to the standard \acrshort{kan} with the same other parts of the workflow in \acrshort{dm}. d) Investigating the decision-making module, we determine the optimal number of clusters for environments with different numbers of known emitters. 

\end{itemize}

This rest of the paper is organized as follows. Sec. \ref{sec:rw} summarizes the related work, Sec. \ref{sec:back} provides the background and problem statement, while Sec. \ref{sec:design} elaborates on the design choices of the emitter detection and identification framework. Sec. \ref{sec:meth} describes the evaluation methodology and Sec. \ref{sec:res} analyzed the results. Finally, Sec. \ref{sec:conc} concludes the paper. 
\section{Related work}
\label{sec:rw}
When focusing on semi-supervised and unsupervised deep learning approaches for detection and identification of emitters the number of related studies is rather low. Table~\ref{tab:combined_sei_ued_summary} summarizes the related works focused on \acrshort{sei} and \acrshort{ued} tasks based on semi-supervised and unsupervised deep learning approaches, structured along the design space presented in Figure ~\ref{fig:design_choices}. As shown, prior works are predominantly focusing on few-shot self-supervised \acrshort{sei} using raw I/Q data with \acrfull{cl} \cite{wu2023specific, zha2023cross, liu2023overcoming, sun2025few} or \acrfull{ae} \cite{huang2022deep, xu2023few, yao2023few, }, with CNN-based feature extractors being the most common backbone. Besides the necessity for labels, the interpretability of these approaches, if supported at all, is limited to the decision-making stage, often via distance-based metrics, while their feature extractors remain opaque.
There is also only one work for self-supervised \acrshort{sei} utilizing 2D constellation data modality \cite{liu2022specific}, which is also designed for few-shot set-up limiting its applications, whereby the authors do not demonstrate either theoretically nor experimentally the need of this transformation.  In this work, we investigate the design space of approaches for self-supervised \acrshort{ued} without using labels during the training, which is a more challenging problem. We prove theoretically and experimentally that in completely label-less case the scenarios of signal transmission greatly affect the correct design choice. We show the advantage of using 2D constellation data modality when transmitted messages are different.    

Despite differences in methodology, only two studies \cite{krasnov2025novel, hao2023contrastive} are fully unsupervised, requiring no labels at any stage. In \cite{hao2023contrastive}, as part of the study, authors demonstrate 91\% accuracy in a zero-shot \acrshort{ued} task using a contrastive learning framework, underscoring the effectiveness of \acrshort{ssl} approaches in overcoming the labeling bottleneck. However, this work uses a proprietary non-public dataset, which makes the results non-replicable. Authors of \cite{krasnov2025novel}, on the other hand, use publicly available WiSig dataset \cite{hanna_wisig_2022}, but it only contains signals transmitting the same messages, making the setup less realistic. They achieve an F1 score between $80\%$ and $87\%$ on self-supervised \acrshort{ued} task depending on the number of emitters in the training set versus all emitters in the testing set. All above-mentioned approaches lack interpretability of the features learning module. 

In this work, we aim to fill the identified gaps by conducting experiments on two publicly available datasets \cite{hanna_wisig_2022, sankhe2019oracle} supporting also the investigation of the impact of transmitting same and different messages. We are focusing on a zero-shot self-supervised open-set \acrshort{ued} task and in contrast to related works we introduce interpretability both at the embedding and feature-extractor levels, providing the first systematic performance-versus-interpretability comparison across \acrfull{dc},  \acrfull{cl}, and \acrfull{ae} paradigms.
\begin{figure}[!t]
\centering
    \includegraphics[width=0.20\textwidth]{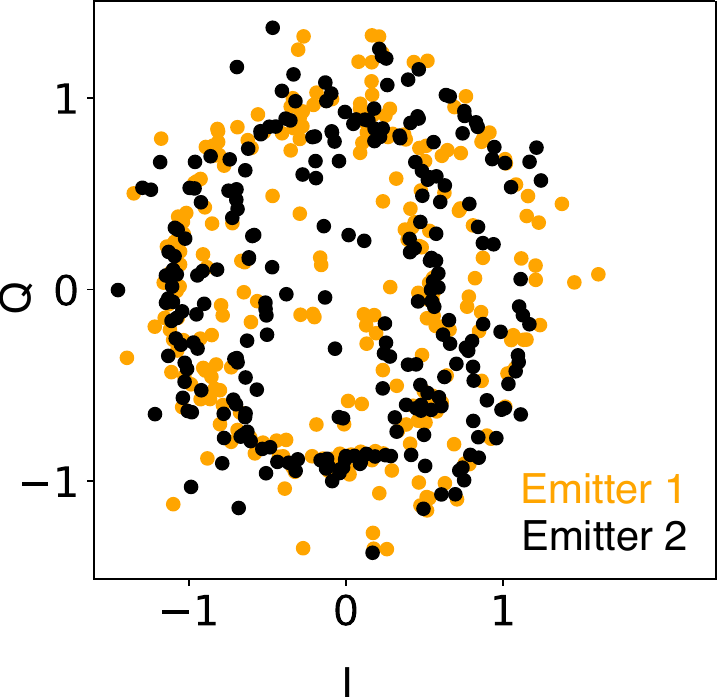}
    \caption{Yellow dots represent an arbitrary signal from the first emitter from the WiSig Dataset \cite{hanna_wisig_2022}, and black dots stand for the arbitrary signal from the second Emitter. The signals carry the same message and differ only by environmental noise and hardware imperfections.}
    \label{fig:wisig_samples_diff}
\end{figure}
\begin{figure*}[!t]
\centering
    \includegraphics[width=\linewidth]{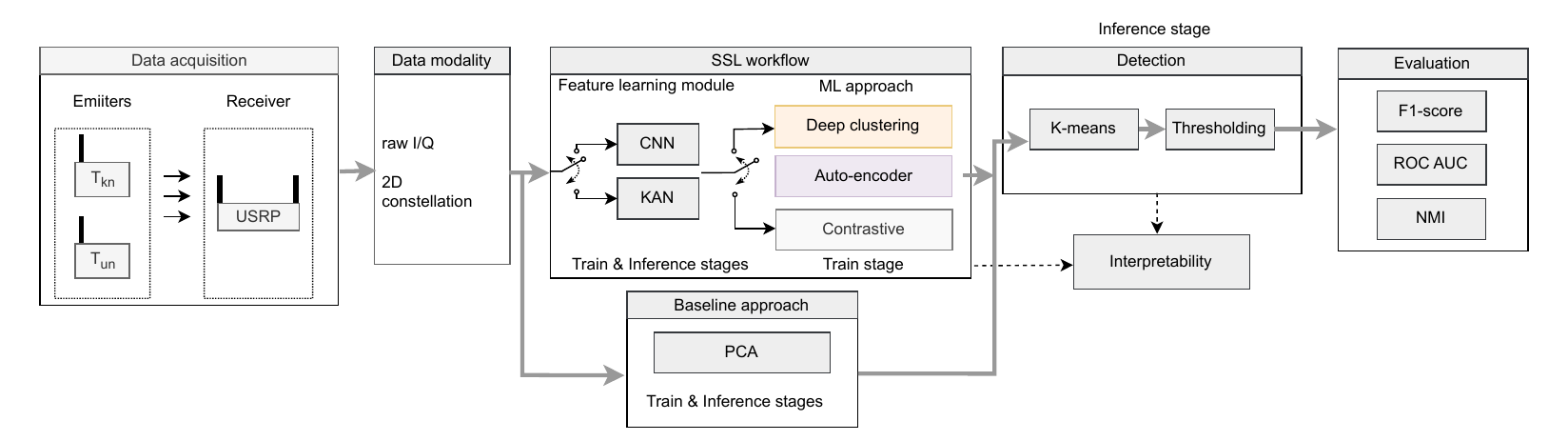}
    \caption{
    Formal self-supervised \acrshort{ml} based \acrshort{ued} workflow.}
    \label{fig:workflow}
\end{figure*}

\section{Background Assumptions and Problem Statement}
\label{sec:back}
In this section, we mathematically describe device-specific signal alterations that are exploited by \acrshort{sei} and \acrshort{ued} tasks and formulate the problem statement of zero-shot self-supervised unknown emitter detection. 
\subsection{RF Fingerprint}
\label{subsec:rf_finger}
In this work, we focus on detection and identification of emitters using Orthogonal Frequency Division Multiplexing (OFDM) modulation scheme. In this modulation scheme, each transmitted symbol corresponds to a complex-valued point in the I/Q (In-phase/Quadrature) constellation diagram, which represents the signal’s amplitude and phase characteristics. Figure \ref{fig:wisig_samples_diff} shows an example constellation diagram of the OFDM modulation scheme with the axes representing the I and Q components of signal points. Each OFDM signal consists of clean modulated message and distortions caused by hardware imperfections and channel noise. While channel noise is affected by environment conditions, hardware imperfections, often referred as radio frequency (RF) fingerprint, are caused by tiny imperfections of individual emitter analog components, making them emitter specific. In this study, we explore these hardware imperfections for distinguishing known and unknown emitters. Simplifying, RF fingerprint \cite{xie2024radio} can be modeled as:
\begin{equation}
x(t) = y(t) \cdot e^{j\left(2\pi(f + \Delta f)t + \phi(t)\right)},
\end{equation}
where $\Delta f$ and $\phi(t)$ are frequency and phase offsets in signal $x(t)$ caused by hardware imperfections. Treating them as small perturbations, we can apply first order Taylor decomposition to divide signal into clean and corrupted components:
\begin{equation}
\label{eq:rf_decomposition}
\begin{split}
&x(t)= y(t) \cdot e^{j2\pi ft + j(2\pi \Delta ft + \phi(t))} \\
&x(t) \approx y(t) \cdot  e^{j 2\pi ft} + y(t) \cdot e^{j 2\pi ft}j(2\pi \Delta ft + \phi(t))\\
&x(t)\approx x_{symbols}(t) + r_{emitter}(t) \\
&|r_{emitter}(t)| \ll |x_{symbols}(t)|
\end{split},
\end{equation}
where $x_{symbols}(t)$ represents clean signal without hardware imperfections and $r_{emitter}(t)$ stands for the RF fingerprint part. Figure \ref{fig:wisig_samples_diff} illustrates the distinctions created by the fingerprint $r_{emitter}(t)$ for two measured transmissions of the same symbols from the WiSig Dataset \cite{hanna_wisig_2022} by two different emitters, where the yellow dots represent Emitter 1 and the black dots represent Emitter 2.   

Clearly, the challenge in distinguishing $r_{emitter}(t)$ in the case of known transmitted messages differs from the case of more general unknown transmitted messages, as we show in this work by considering two datasets. The physical (PHY) layer of transceiver is responsible for sending the data over the wireless medium by modulating it onto a carrier signal. It consists of a preamble, a header and a payload. The preamble and  header have fixed length, whereby the first also has a fixed repeating sequence of bits used for synchronization while the second contains varying control information like the transmission rate and payload length. These are followed by a varying payload of the length specified in a header containing a medium access control (MAC) frame with the actual data and a trailer with error-checking information. Taking this into account, we consider in this work two signal receiving scenarios: \textit{Same Messages  Scenario} for the cases when only the preamble symbols are considered and \textit{Different Messages Scenario} when all received symbols are considered. 
\subsection{\acrfull{sm}}
\label{subsec:preamle_scen}
If only looking at a preamble, a signal with the same content is transmitted by every emitter in the network. The preamble signal coming from the $i^\textbf{th}$ emitter, using Eq.~\ref{eq:rf_decomposition}, can be described as:
\begin{equation}
    \label{eq:rf_samemsgs}
    x_{i}(t) =x_{symbols}(t) + r_{emitter,i}(t),
\end{equation}
where $x_{symbols}(t)$ is the constant preamble signal, and $r_{emitter,i}$ is the $i^\textbf{th}$ emitter specific signal. Then, the difference between the preamble signals from different emitters is higher than the variation inside each emitter:
\begin{equation}
    \begin{split}
        &x_{i}(t) -x'_i(t) \approx0 \\
        &x_i(t)-x_j(t)=r_{emitter,i}(t) -r_{emitter,j}(t) \neq0 .
    \end{split}
\end{equation} 
This makes the self-supervised task easier because of linearly separable classes.
\subsection{\acrfull{dm}}
\label{subsec:random_scen}
In a more generic use case when we do not know the exact technology and protocol used, we may only have access to some random part of the received frame, which means that the clean signal $x_{symbols}$ is also unknown. The signal can now be expressed as:
\begin{equation}
    \label{eq:rf_diffmsgs}
    x_{i}(t) =\hat{x}_{symbols}(t) + r_{emitter,i}(t),
\end{equation}
where $\hat{x}_{symbols}(t)$ is random. Thus, the variation within the emitter becomes the same as differences between emitters because the emitter specific part represents a small perturbation compared to the clean symbol as stated in Eq. \ref{eq:rf_decomposition} and expressed as follows:
\begin{equation}
    \begin{split}
        &x_{i}(t) -x'_i(t) \approx \hat{x}_{symbols}(t)-\hat{x}'_{symbols}(t) \\
        &x_i(t)-x_j(t)\approx \hat{x}_{symbols}(t)-\hat{x}''_{symbols}(t)
    \end{split},
\end{equation}
where $\hat{x}_{symbols}(t), \hat{x}'_{symbols}(t), \hat{x}''_{symbols}(t)$ are random signals. This more reasonable assumption makes self-supervised task much harder.   

\subsection{Problem Statement}
\label{sec:problem_rf}

In this work, we formulate the problem of \acrshort{ued} as a \textit{zero-shot learning task} within a \textit{self-supervised learning setting}. We develop an approach that extracts the RF fingerprint part from signals, as shown in Eq. \ref{eq:rf_decomposition}, grouping identical emitters and separating different ones in the feature space. The objective is to learn a composite function
\begin{equation} \label{eq:composite_function}
\Phi : X_{\text{in}} \rightarrow L, \quad \text{where } L = \{\text{known}\: (0), \text{unknown}\:(1)\}
\end{equation}
\noindent that maps raw RF signal samples into corresponding decisions indicating whether a signal originates from a previously seen or an unseen emitter. This is done without access to ground-truth labels during training and under the constraint that the number of emitters in the training set is not known apriori.

The composite function $\Phi(X_{\text{in}})$ from Eq.~(\ref{eq:composite_function}) can be decomposed into two modular sub-functions as per Eq.~(\ref{eq:modular_decision}): the embedding function \( F \), which transforms raw RF data into a projection (i.e. embedding) in a discriminative feature space, and the detection function \( M \), which determines whether an embedded sample belongs to a known or an unknown emitter. 
\begin{equation}
\label{eq:modular_decision}
\text{L} = \Phi(X_{\text{in}}) = M(F(X_{\text{in}}))
\end{equation}
\subsubsection{Embedding Function} \label{sec:embedding_func_rf}
Let \( F \) be the embedding function trained in a self-supervised manner to project input signals into a latent space $
F : X_{\text{in}} \rightarrow \mathbb{R}^d$,
where \( X_{\text{in}} \) denotes the input signal domain, and \( \mathbb{R}^d \) is the latent feature space of dimension \( d \). The output embeddings should cluster tightly for samples from the same emitter while remaining well-separated for different emitters. In other words, the latent space has to represent fingerprint components of signals from Eq. \ref{eq:rf_decomposition}.

\subsubsection{Detection Function} \label{sec:detection_func_rf}

The detection function \( M \) operates over the embedded space to determine whether a given sample is associated with a known or unknown emitter $M : \mathbb{R}^d \rightarrow L$.

\section{Design of the \acrshort{ued} workflow}
\label{sec:design}
To address the problem identified in Sec. \ref{sec:problem_rf}, we formalize the structured workflow for ML-based \acrshort{ued} as depicted in Figure \ref{fig:workflow}. The most important components of this workflow are: a) data modality, b) feature learning module (\acrshort{flm}), c) machine learning approach and d) decision making module. Data modality component transforms the input signal to the chosen representation: raw I/Q or 2D constellation. The \acrshort{flm} and machine learning approach represent the self-supervised learning block. The selected \acrshort{flm}: KAN or CNN is trained by one of the machine learning approaches: \acrshort{dc},\acrshort{cl}, \acrshort{ae} in a self-supervised manner. Principal component analysis (PCA) is a prominent baseline in unsupervised learning, but it does not fit within a feature learning model-\acrshort{ml} approach scheme. Thus, we treat it as an individual feature-learning module with a specialized training procedure. The decision making module is the final stage of the \acrshort{ml} workflow, where a classification decision is made based on extracted features. The other components include evaluating metrics and interpreting the \acrshort{flm} and \acrshort{dmm}. While not being directly involved, they are essential for the performance measuring and analysis of the system. The following subsections provide a detailed design description of each component.
\subsection{Transmitting devices}\
\label{subsec:same_and_diff_scenraious}
In this study, devices transmit signals following two scenarios: the \acrlong{sm} as described in Sec. \ref{subsec:preamle_scen} and \acrlong{dm} as per Sec. \ref{subsec:random_scen}. The first scenario, in which all emitters transmit the same content all the time, is considered in the case where one has access to the preamble. The second scenario with different content is more realistic but also more challenging. The  particular scenario affects the best design choices of Unknown Emitter self-supervised zero-shot detectors. 
\subsection{Data Modality}
\begin{figure}[!t]
\centering
\begin{subfigure}[b]{0.27\textwidth}
\includegraphics[width=\textwidth]{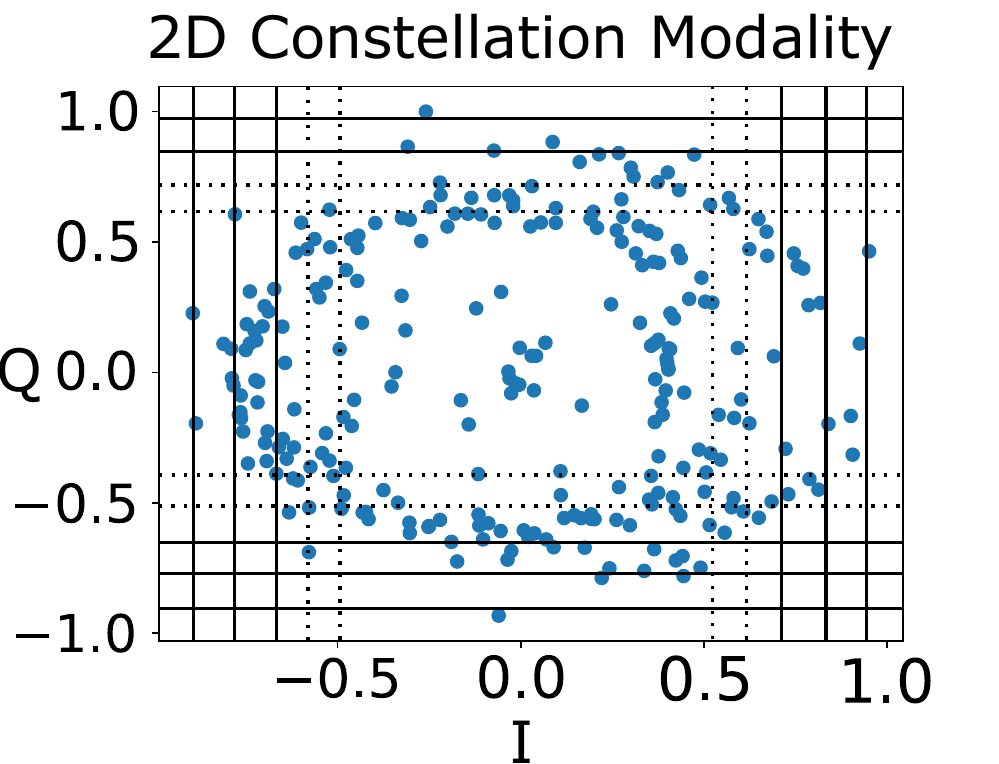}
    \caption{2D constellation Modality construction from Raw I/Q with Grid Size $K=60$}
    \label{fig:2D_const_grid}
\end{subfigure}
\hfill
\centering
\begin{subfigure}[b]{0.21\textwidth}
\includegraphics[width=\textwidth]{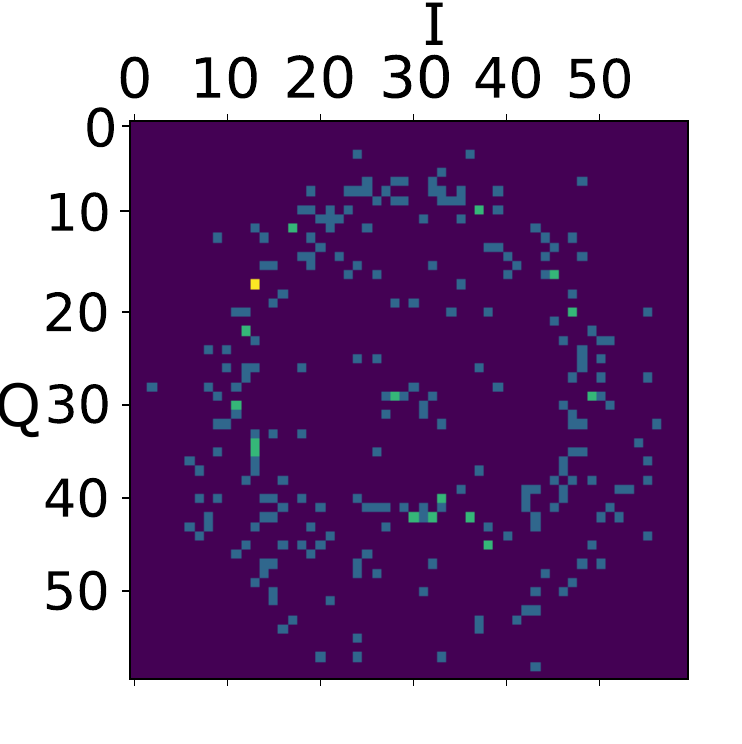}
    \caption{2D constellation Modality Single Channel Image with Grid Size $K=60$}
    \label{fig:2D_const_final}
\end{subfigure}
\caption{2D constellation Modality Description}
\label{fig:2D_const}
\end{figure}
In the receiver, the RF signal is discretized with the same time step:
$x[i] = x(i\delta t)$, where $x[i]$ is a complex value denoting I and Q components. The way it is processed to be fed to the features learning module $F$ is referred to as a data modality, an important part of the design space. In this work, we investigate two types of data modalities: raw I/Q data and a 2D constellation. The raw I/Q is the convenient choice for OFDM data; each observation \( x[i] = x[i]^I + jx[i]^Q \in \mathbb{C}^N \) is represented as a two-channel input:
$X_{\text{in}}^{0i} = x^I[i], \quad X_{\text{in}}^{1i} = x^Q[i]$.
 In this study, we argue that the raw I/Q data modality produces poor quality under the realistic scenario, assuming different transmitting messages as described in Sec. \ref{subsec:random_scen} because of the prevalence of semantic information $x_{symbols}(t)$ over the fingerprint information $r_{emittter}(t)$ according to Eq. \ref{eq:rf_diffmsgs}. To address this issue, one can transform the data in a time-invariant manner:
\begin{equation}
    \label{eq:time_inv}
    T(X_{\text{in}}) = T(\sigma(X_{\text{in}})), \sigma(X_{\text{in}})^{ij} = X^{i,\sigma(j)}_{\text{in}},
\end{equation}
where $T$ is a transformation operation over raw I/Q data, and $\sigma$ is a permutation operation over the time axis. 
The \(T(X_{\text{in}})\) tensor does not contain any semantic information, because all messages have the same representation, assuming that transmitted signals are random. A straightforward way to provide the time-invariant property is to represent raw I/Q samples as a 2D constellation data. Firstly, the data is scaled to $[-1,1]$ range:
\begin{equation}
    \label{eq:Xin_scaling}
    X_{\text{in norm}}= \frac{X_{in}}{max(|X_{in}|)}.
\end{equation}

Secondly, the procedure counting I/Q observations over cells created by the grid on a 2D plane is applied as depicted in the Figure \ref{fig:2D_const_grid}: 
\begin{equation}
\label{eq:2d_const}
    \begin{split}
    T(X_{\text{in}})^{i,j} = \#\{k: \frac{X_{\text{in norm}}^{0k} + 1}{2} \in [i\epsilon,(i+1)\epsilon],\\ \frac{X_{\text{in norm}}^{1k} + 1}{2} \in [j\epsilon,(j+1)\epsilon]\} / N
    \\
    \text{where } i,j \in[0,K-1], \epsilon=\frac{1}{K},
    \end{split}, 
\end{equation} 
\noindent where $K$ is the grid size and $N$ is the number of I/Q observations. The resulting tensor looks like a heat map shown in Figure~\ref{fig:2D_const_final}.

\subsection{\acrlong{flm} }
\label{susbsec:flm}
The \acrshort{flm} $F$ is a core part of self-supervised zero-shot unknown emitter detectors. Based on the available literature \cite{liu2023overcoming, hao2023contrastive}, Convolutional Neural Networks (CNNs) based architectures are employed for \acrshort{flm} s. To provide a more in-depth exploration of design principles of this component, we also study  new, interpretable \acrshort{kan} based extractors and PCA baselines.

With respect to CNNs, we investigate 1D and 2D architectures for raw I/Q and 2D constellation data modalities, respectively. The CNN-1D two-channel network, adopted from \cite{o2018over}, consists of four blocks, each containing convolution, batch normalization, and max pooling operations. 2D CNN is similar to CNN-1D but with 2D spatial operations. However, CNN architectures are not interpretable, as their decisions are not understandable by humans. 

As a novel, interpretable type of architecture, we investigate Kolmogorov-Arnold Networks (\acrshort{kan}s) \cite{liu2024kan} in the design space to bring interpretability to unknown emitter detectors. The single \acrshort{kan} layer can be described as an MLP layer, where activations are replaced with learnable functions and fixed weights on the edges. The learnable activation functions are often drawn from the k-order spline functions \cite{de1978practical}, which can be parameterized as linear combinations of basic B-splines. These linear coefficients are then optimized during the training process. We develop a single-layer Efficient \acrshort{kan} version as:
\begin{equation}
    F^{(j)}(x_1, ..x_N) = \sum_i^N \psi_{ij}(x_i),
\end{equation} 
where $j \in[1,N^*]$ is a coordinate of the feature space, $(x_1, ..., x_N)$ is an input tensor $X_{in}$ divided into individual nodes, and $\psi_{ij}$ are functions which are represented as:
\begin{equation}
    \psi_{ij}(x) = w_{ij}^{(b)}silu(x)+w_{ij}^{(s)}\sum_{k=1}^{G+4}\gamma_{i,j}^{(k)}\phi_{i,j}^{(k)}(x),
\end{equation}
where $\phi_{i,j}^{(k)}(x)$ are fourth-order B-splines, $silu$ is Sigmoid Linear Unit (SiLU) activation function and $G$ is a grid size.  

For CNNs and \acrshort{kan}s, we also investigate SVD initialized modifications of those architectures. Let $F$ be a \acrshort{flm} (CNN or \acrshort{kan}), which transforms the input data to $d$-dimensional features space. We initialize a linear layer with SVD weights with the number of latent components $d$ and combine this linear layer with the initial \acrshort{flm} $F$: 
\begin{equation}
    F'(x) = \frac{F(x)}{10} + L(x),
\end{equation}
where $L$ is the linear layer initialized with SVD weights. Under conditions of sparse and limited data, like in 2D constellation data modality, this modification aims to provide additional knowledge of the linear nature of the RF fingerprint (see Eq. \ref{eq:rf_decomposition}) and boost the performance of a complex self-supervised approach, which might degrade under the conditions as mentioned earlier \cite{arora2017simple, fournier2019empirical}. 
This transformation does not move the \acrshort{kan} model out of \acrshort{kan} architectures, because adding linear functions to cubic splines produces cubic splines. 

We also investigate PCA - a prominent baseline in unsupervised learning. It builds a linear transformation that extracts low-dimensional, most informative features based on their contributions to total variation. 

\subsection{Machine Learning Approach}
\label{sec:mlapproach}
According to Sec. \ref{sec:rw}, there are three main \acrshort{ml} approaches used in self-supervised \acrshort{sei}/\acrshort{ued}: \acrlong{dc}, \acrlong{ae}, and \acrlong{cl}. These approaches are used to train the \acrshort{flm} and are typically followed by a decision making module at inference time. In this paper, we investigate the design dimension of the \acrshort{ml} approach by evaluating the performance trade-offs of the three selected approaches. 

\subsubsection{Deep Clustering}
\acrshort{dc} \cite{Caron_2018_ECCV} is an \acrshort{ml} approach in self-supervised learning, which is employed in \cite{krasnov2025novel}. The training procedure consists of two alternating stages. During the first stage, features are extracted from the training set using the current state of \acrshort{flm} and clustered using the K-means algorithm, assigning cluster numbers (pseudo-labels) to training samples. At the second stage \acrshort{flm} is updated using the classification objective with those pseudo-labels. The clustering results with label assignment are used solely for the \acrshort{flm} $F$'s training objective and are not reused thereafter for inference or decision making.
\subsubsection{Auto Encoders}
\acrshort{ae} \cite{liu2021self} is a classical  self-supervised \acrshort{ml} approach, used in \cite{huang2022deep}. It learns the \acrshort{flm} through introducing a decoder network and a reconstruction objective.
The loss function used is the mean square error (MSE) between the input and its reconstruction. In the \acrshort{ae} approach, learnable decoders are required. We use CNN with deconvolutions as the decoder for the CNN architecture and the \acrshort{kan} layer as the decoder for the \acrshort{kan} architecture. 
\subsubsection{Contrastive learning}
We investigate the SimCLR \cite{chen2020simple} \acrshort{cl} approach similar to \cite{hao2023contrastive}, while \cite{sun2025few, liu2023overcoming, wu2023specific} also employ \acrshort{cl} but in slightly different set-ups.
For each sample, two augmented views are generated via random transformations. Views of the sample are treated as positive pairs, and others are treated as negative pairs.
Features are extracted from augmented samples using \acrshort{flm}, and a cross-entropy-like loss  computed with objective to classify positive and negative pairs.   
\begin{figure}[!t]
\centering
\includegraphics[width=0.43\columnwidth]{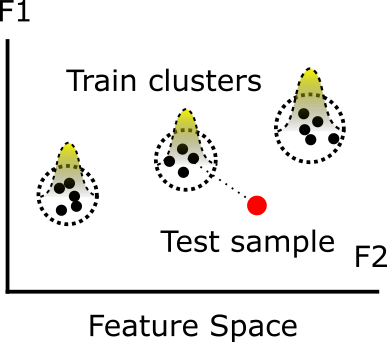}
    \caption{Decision Making on a test sample. Three circles represent training clusters with probability densities equal to zero outside the circle. The red point represents a test sample, which, in this case, is an outlier and is predicted as \textit{unknown}. $F1,F2$ are axes of demonstrative 2D features space.}
    \label{fig:dmm}
\end{figure}
\subsection{Decision Making Module}
\label{subsec:dmm}
\acrshort{dmm} is the final stage of \acrshort{ml} based \acrshort{ued} as depicted in Figure \ref{fig:workflow}. It has to be independent of the choice of a specific \acrshort{flm} or \acrshort{ml} approach. There are several widely used \acrshort{dmm}s like K-means based \cite{smiti2020critical}, KNN based \cite{hao2023contrastive}, MLP-based \cite{krasnov2025novel}, etc. As we solve the \acrshort{ssl} problem of outliers detection, we investigate the cluster-based \acrshort{dmm} \cite{smiti2020critical}, which is a robust, universal, and interpretable method for outlier detection. After the training, we extract the training features using a trained \acrshort{flm} $F$ and cluster them into $C$ clusters with the K-means algorithm. During inference, we extract features from the test sample and identify the nearest cluster. Then, we compare the distance from the test sample to the nearest cluster center with the distances of the train samples in this cluster. If a test sample lies in the right $\alpha 
\% $ quantile of the train distribution, it is marked as unknown. This concept is illustrated in Figure~\ref{fig:dmm}. 

Formally, if $d_i$ is a distance from the test sample to the nearest cluster center and $\{d'_j\}_{j=1}^{N_{tr}}$ are distances from the same cluster center to train samples lying in it, then the classification score $s_i$ is computed as:
\begin{equation} \label{eq:s_score}
    s_i = \frac{\#\{j:d_i>d_j\}}{N_{tr}}.
\end{equation}
If $s_i > \alpha$, then the test sample is labeled as unknown. Classification scores from Eq.~(\ref{eq:s_score}) and the indices of nearest clusters are further used for metric computation. 

\section{Methodology}
\label{sec:meth}
In this section, we describe the datasets used for evaluation, followed by the evaluation metrics and experiment setups.
\subsection{Datasets}
As mentioned in Sec.~\ref{subsec:same_and_diff_scenraious}, we leverage the \acrshort{ml}-based  \acrshort{ued} workflow under two scenarios to study the design principles of zero-shot self-supervised detection. For the study we rely on the WiSig \cite{hanna_wisig_2022} and ORACLE \cite{sankhe2019oracle}  datasets designed for the same and different messages scenarios, respectively, sharing similar indoor environment conditions. 
\begin{figure}[!t]
\centering
\includegraphics[width=0.9 \columnwidth]{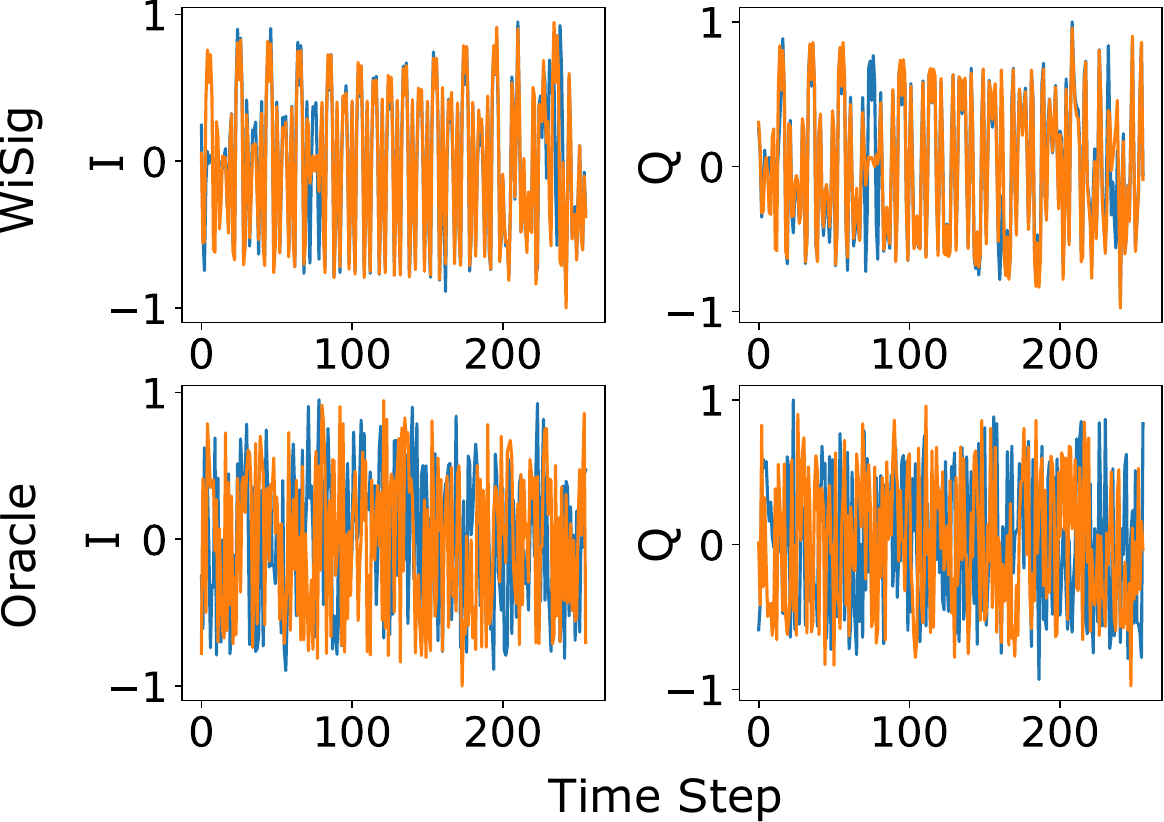}
    \caption{WiSig and ORACLE datasets comparison. The first column shows I components, and the second column represents Q components of the I/Q samples. Different colors refer to different signal traces from the datasets.}
    \label{fig:same_diff_scen}
\end{figure}
\subsubsection{WiSig}
\label{subsec:wisig_descr}
The WiSig dataset specifically includes signals captured from six distinct WiFi transmitting devices, recorded over a period of four consecutive days in indoor conditions. For each device, $1,000$ signals were recorded per day, resulting in a well-structured dataset designed to facilitate emitter identification tasks. Each signal trace consists of 256 consecutive I/Q samples, capturing a short time window of the emitted waveform. The upper line in Figure~\ref{fig:same_diff_scen} shows the I and Q components of two signal traces from the WiSig dataset; the repeating patterns indicate that the messages of the two signal traces are the same.
\subsubsection{ORACLE}
\label{subsec:oracle_descr}
The ORACLE dataset was also collected in indoor conditions. It consists of 16 emitters, each containing two long signals of 40 million I/Q observations recorded at a distance of 20 meters. Those signals are sliced into short traces of 256 I/Q samples each, similar to those in the WiSig dataset. The number of samples per emitter is reduced to match the WiSig dataset, i.e., $4,000$. The bottom line in Figure~\ref{fig:same_diff_scen} shows I and Q components of two signal traces from the ORACLE dataset; different patterns indicate that the messages of two signal traces are different.

\subsection{Metrics}
For evaluation in terms of clustering quality and detection performance, we use three complementary metrics, i.e. 
the area under the Receiver Operating Characteristic curve (ROC-AUC), Normalized Mutual Information (NMI), and  F1-score. We calculate these metrics using predictions from \acrshort{dmm} described in Sec.~\ref{subsec:dmm}.

The \textit{ROC-AUC} metric measures the ability of the model to discriminate between classes across varying decision thresholds. 
A higher AUC value indicates better separability between known and unknown samples. 
Unlike F1, ROC-AUC provides a threshold-independent evaluation of detection quality. For ROC-AUC calculation, classification scores from \acrshort{dmm} are used with $\{known  (0), unknown(1)\}$ labels in the test dataset. 

The \textit{NMI} metric quantifies the agreement between predicted cluster assignments and ground-truth labels.
It measures how much information is shared between the predicted and actual label distributions, 
where $0$ indicates no mutual information and $1$ denotes a perfect clustering alignment. NMI is calculated between assigned clusters' indices from \acrshort{dmm} and emitters' IDs.

\textit{F1-score} metric represents the harmonic mean between precision and recall, 
defined as $F1 = 2 \cdot \frac{\text{Precision} \cdot \text{Recall}}{\text{Precision} + \text{Recall}}$. 
It balances false positives and false negatives, making it an effective indicator of detection reliability in both closed- and open-set conditions. In our context, the F1-score is computed for the binary classification task of distinguishing between known and unknown samples. We obtain predictions via the procedure from \acrshort{dmm} with $\alpha=5\%$. Along with performance metrics, the numbers of learnable parameters and floating point operations (FLOPs) are calculated for each \acrshort{flm}.  

\subsection{Experiments}
In the following, we outline the experiment setups for the two scenarios, i.e., with all emitters transmitting the same messages or different messages.

\subsubsection{\acrfull{sm}}
\label{subsec:meth_same}
For the \acrshort{sm} scenario described in Sec. \ref{subsec:preamle_scen}, we employ the WiSig dataset described in Sec.~\ref{subsec:wisig_descr}. 
We evaluate both CNN-1D and \acrshort{kan} architectures, each producing feature embeddings of size 20, with DMM configured to use 80 clusters. 
The four-layer CNN-1D model follows the architecture proposed in \cite{o2018over}, while the \acrshort{kan} model is implemented with a grid size of 10, input dimensionality of 512, and spline order of 4. 
For the \acrshort{ae} paradigm, we employ a 1D CNN-based encoder and decoder. The former is a convenient CNN model with reducing image size and increasing number of channels, and the latter is constructed by replacing convolutional layers with deconvolutional layers and max-pooling operations with up-sampling, following the design in \cite{chen2020one}. 
The \acrshort{kan}-based \acrshort{ae} uses a single-layer \acrshort{kan} architecture comprising 20 input nodes and 512 output nodes. 
In the Contrastive Learning framework, Gaussian noise with a mean of zero and a standard deviation of 0.05 is added to input samples as a data augmentation technique along with random rotations and amplifying similarity to \cite{hao2023contrastive}.

For evaluation, we employ a cross-validation procedure in which, iteratively, one of the six emitters is designated as the unknown test emitter. 
In each iteration, the training dataset consists of signals from the remaining five emitters, collected across all days and using the first 80\% of samples from each day. 
The test dataset includes data from all six emitters, assembled from the remaining $20\%$ of samples per day. 
Within this test set, the selected emitter is labeled as \textit{unknown}, while the other five are treated as \textit{known}. 

Both the CNN-1D and \acrshort{kan} architectures are trained for 100 epochs on the training dataset under three learning paradigms: Deep Clustering, Auto-Encoding, and Contrastive Learning with a learning rate $10^{-3}$ and an Adam optimizer. 
During training, evaluation metrics are computed on the current test dataset every 15 epochs. 
For each evaluation step, metrics are averaged across the six cross-validation iterations. 
The epoch corresponding to the maximal average performance is then selected, and the metrics at this epoch are reported as final results. 
This approach follows the standard practice in unsupervised and open-set learning, where the final epoch may not represent the best model state due to non-monotonic training dynamics \cite{caron2018deep, chen2020simple}. 
A similar evaluation strategy was previously adopted in \cite{jain2014multi}. 

Additionally, we evaluate the PCA with feature dimensionality of 20, varying the number of clusters in \acrshort{dmm} among $\{10, 40, 80, 120, 160, 200, 240\}$, as well as \acrshort{kan}-based models with feature sizes $\{2, 3, 5, 10, 20\}$, using the same cross-validation procedure.
\subsubsection{\acrfull{dm}}
For the \acrshort{dm} scenario described in Sec. \ref{subsec:random_scen}, we use the ORACLE dataset described in Sec.~\ref{subsec:oracle_descr}. We perform the experiments using two data modalities: raw I/Q data and 2D constellation data (Figure \ref{fig:2D_const} and Eq. \ref{eq:2d_const}) with a grid size of $G=60$.
The approaches and models for raw I/Q modulation are the same as for the \acrshort{sm} as explained in Sec. \ref{subsec:meth_same}. For the 2D constellation modality, we evaluate a four-layer 2D CNN, structurally similar to the previously described CNN-1D variant, along with \acrshort{kan}s and their SVD-initialized counterparts from Sec. \ref{susbsec:flm}. For the \acrshort{ae} approach, a 2D CNN decoder is utilized, which is also similar to a CNN-1D Decoder.
All feature extractors produce embeddings of size 20, and DMM operates with 200 clusters. 
For \acrshort{kan}-based architectures, the input layer consists of 3600 nodes. 
In the contrastive learning setup, data augmentations include rotations of $0$, $\pi$, $\frac{\pi}{2}$, and $\frac{3\pi}{2}$ radians, as well as additive Gaussian noise with a standard deviation of $0.05$. 
Each \acrshort{flm} is evaluated under three learning paradigms: Deep Clustering, Auto-Encoding, and Contrastive Learning. We also run the PCA method with feature size 20 across $\{10, 40, 80, 120, 160, 200, 240\}$ numbers of clusters in \acrshort{dmm}. For the evaluation, we use a scheme similar to Sec. \ref{subsec:meth_same}. The only difference is that we randomly shuffle the emitters beforehand and consequently choose two emitters as unknown, making a total of five folds.

\begin{table*}[h!]
\caption{\acrshort{sm}: WiSig dataset results with the raw I/Q data modality.}
\label{table:wisig_perf}
\centering
\begin{tabular}{|c|c|c|c|c|c|c|c|c|c|}
\hline
\textbf{Approach} & \textbf{FE Module} & \textbf{Modality} & \textbf{F.Size} & \textbf{Clusters} & \textbf{ROC-AUC \%} & \textbf{NMI \%} & \textbf{F1 \%} & \textbf{Params} & \textbf{FLOPs} \\
\hline
\acrshort{dc} & CNN-1D  & raw I/Q & 20 & 80 & 95.4$\pm$1.9 & 51.5$\pm$0.8 & 76.0$\pm$6.2 & 80K & 1M \\
\acrshort{dc} & \acrshort{kan} & raw I/Q & 20 & 80 & 97.0$\pm$2.5 & 55.4$\pm$0.9 & 77.5$\pm$8.7 & 143K & 6M \\

\acrshort{ae} & CNN-1D  & raw I/Q & 20 & 80 & 99.2$\pm$0.5 & 57.4$\pm$0.7 & 87.6$\pm$1.2 & 80K & 1M \\
\acrshort{ae} & \acrshort{kan} & raw I/Q & 20 & 80 & 99.3$\pm$0.7 & 58.9$\pm$0.8 & 88.0$\pm$1.3 & 143K & 6M \\

\acrshort{cl}& CNN-1D  & raw I/Q & 20 & 80 & 97.6$\pm$1.1 & 51.4$\pm$1.1 & 82.7$\pm$4.9 & 80K & 1M \\
\acrshort{cl}& \acrshort{kan} & raw I/Q & 20 & 80 & 97.8$\pm$1.3 & 55.9$\pm$1.7 & 83.8$\pm$3.8 & 143K & 6M \\

PCA & PCA & raw I/Q & 20 & 80 & 98.7$\pm$1.1 & 56.2$\pm$0.9 & 84.8$\pm$4.1 & 10k & 0.2M \\
\hline
\end{tabular}
\end{table*}

\begin{table*}[h!]
\caption{\acrshort{dm}: ORACLE dataset results with the raw I/Q data modality.}
\label{table:oracle_perf}
\centering
\begin{tabular}{|c|c|c|c|c|c|c|c|c|c|}
\hline
\textbf{Approach} & \textbf{FE Module} & \textbf{Modality} & \textbf{F.Size} & \textbf{Clusters} & \textbf{ROC-AUC \%} & \textbf{NMI \%} & \textbf{F1 \%} & \textbf{Params} & \textbf{FLOPs} \\
\hline
\acrshort{dc} & CNN-1D  & raw I/Q & 20 & 200 & 50.1$\pm$2.2 & 2.4$\pm$0.8 & 10.1$\pm$1.2 & 80K & 1M \\
\acrshort{dc} & \acrshort{kan} & raw I/Q & 20 & 200 & 49.5$\pm$0.9 & 0.5$\pm$0.1 & 9.1$\pm$0.6 & 143K & 6M \\
\acrshort{ae} & CNN-1D  & raw I/Q & 20 & 200 & 50.7$\pm$3 & 9.4$\pm$5.4 & 9.8$\pm$3.3 & 80K & 1M \\
\acrshort{ae} & \acrshort{kan} & raw I/Q & 20 & 200 & 50.1$\pm$0.6 &  2.1$\pm$0.7 & 10.2$\pm$0.7 & 143K & 6M \\
\acrshort{cl}& CNN-1D  & raw I/Q & 20 & 200 & 51.3$\pm$1.7 & 10.6$\pm$1.5 & 12.1$\pm$2.7 & 80K & 1M \\
\acrshort{cl}& \acrshort{kan} & raw I/Q & 20 & 200 & 49.4$\pm$0.5 & 1.2$\pm$0.6 & 10.1$\pm$0.5 & 143K & 6M \\
PCA & PCA & raw I/Q & 20 & 200 & 42.1$\pm$2.3 & 4.9$\pm$0.3 & 3.4$\pm$1.3 & 10k & 0.2M \\
\hline
\end{tabular}
\end{table*}

\definecolor{high}{RGB}{144,238,144} 
\definecolor{mid}{RGB}{255,255,200} 
\definecolor{low}{RGB}{255,182,193} 

\begin{table*}[h!]
\caption{\acrshort{dm}: ORACLE dataset results with the 2D constellation data modality. Rows shaded according to ROC-AUC score (green = best, yellow = medium, unshaded = low).}
\label{table:oracle_perf_const}
\centering
\begin{tabular}{|c|c|c|c|c|c|c|c|c|c|}
\hline
\textbf{Approach} & \textbf{FE Module} & \textbf{Modality} & \textbf{F.Size} & \textbf{Clusters} & \textbf{ROC-AUC \%} & \textbf{NMI \%} & \textbf{F1 \%} & \textbf{Params} & \textbf{FLOPs} \\
\hline

\rowcolor{high}
\acrshort{dc} & \acrshort{kan} SVD Init & 2D constellation & 20 & 200 & 71.5$\pm$5.1 & 48.9$\pm$1.7& 41.1$\pm$8.1 & 720K & 1.8M \\ 

\acrshort{ae} & \acrshort{kan} SVD Init & 2D constellation & 20 & 200 & 52.3$\pm$2.9 & 21.9$\pm$1.3 & 10.3$\pm$2.1 & 720K & 1.8M \\ 

\rowcolor{mid}
\acrshort{cl}& \acrshort{kan} SVD Init & 2D constellation & 20 & 200 & 64.2$\pm$6.2 & 33.7$\pm$0.8 & 26.7$\pm$9.7 & 720K & 1.8M \\ 

\rowcolor{mid}
\acrshort{dc} & CNN-2D SVD Init & 2D constellation & 20 & 200 & 65.8$\pm$6.5 & 43.3$\pm$2.7 &  24.1$\pm$6.1 & 180K & 11M \\ 

\acrshort{ae} & CNN-2D SVD Init & 2D constellation & 20 & 200 & 51.1$\pm$3.2 & 0.5$\pm$0.1 & 15.6$\pm$2.1 & 180K & 11M \\ 

\rowcolor{mid}
\acrshort{cl}& CNN-2D SVD Init & 2D constellation & 20 & 200 & 66.2$\pm$6.4 & 34.2$\pm$0.9 & 29.5$\pm$9.6 & 180K & 11M \\ 
\hline \hline

\acrshort{dc} & \acrshort{kan} & 2D constellation & 20 & 200 & 48.5$\pm$0.09 & 5.3 $\pm$0.5 &  8.7$\pm$5.3 & 720K & 1.8M \\ 
\acrshort{ae} & \acrshort{kan} & 2D constellation & 20 & 200 & 49.7$\pm$1.2 & 6.1$\pm$0.6 & 11.9$\pm$1.6 & 720K & 1.8M \\ 
\acrshort{cl}& \acrshort{kan} & 2D constellation & 20 & 200 & 57.1$\pm$4.5 & 34.9$\pm$1.4 & 15.8$\pm$5.4 & 720K & 1.8M \\ 

\acrshort{dc} & CNN-2D & 2D constellation & 20 & 200 & 49.6$\pm$1.1 & 2.4 $\pm$0.6 &  9.7$\pm$1.2 & 180K & 11M \\ 
\acrshort{ae} & CNN-2D & 2D constellation & 20 & 200 & 49.7$\pm$1.2 & 6.1$\pm$0.6 & 11.9$\pm$1.6 & 180K & 11M \\ 

\rowcolor{mid}
\acrshort{cl}& CNN-2D & 2D constellation & 20 & 200 & 66.5$\pm$8.2 & 33.6$\pm$0.9 & 29.7$\pm$9.5 & 180K & 11M \\ 

PCA & PCA & 2D constellation & 20 & 200 & 60.1$\pm$5.2 & 34.3$\pm$0.8 & 17.1$\pm$0.6 & 72k & 0.1M \\
\hline
\end{tabular}
\end{table*}

\section{Results}
\label{sec:res}
In this section, we analyze aspects of designing the various components of the workflow illustrated in Figure \ref{fig:workflow}. The results are organized as a discussion of the design choices illustrated in Figure~\ref{fig:design_choices}. Table \ref{table:wisig_perf} summarizes the performance of \acrshort{kan} and CNN architectures across three self-supervised learning paradigms. The \textit{Approach} and \textit{FE Module} columns specify the \acrshort{ml} approach discussed in Sec. \ref{sec:mlapproach} and \acrshort{flm} from Sec. \ref{susbsec:flm}, respectively. The columns \textit{ROC-AUC}, \textit{NMI}, and \textit{F1} report the corresponding evaluation metrics (in percentages), whereas \textit{Params} and \textit{FLOPs} columns indicate the number of parameters and the number of floating-point operations required to process a single sample.  Tables \ref{table:oracle_perf} and \ref{table:oracle_perf_const} follow the same organization as Table \ref{table:wisig_perf}, but report results for the ORACLE dataset with different data modalities.  

\subsection{The impact of data modality}
\label{subsec:results_data_modality}

As can be seen from Tables \ref{table:wisig_perf}, \ref{table:oracle_perf}, and \ref{table:oracle_perf_const}, raw I/Q data modality shows high performance in the \acrshort{sm} scenario. It demonstrates severely degraded performance in the \acrshort{dm} scenario, whereas using the 2D constellation data modality significantly improves performance in the \acrshort{dm} setup due to the time-invariant representation property described in Sec.~\ref{sec:design}. 

As shown in Table~\ref{table:wisig_perf} for the \acrshort{sm} scenario, \textit{PCA, \acrshort{kan} trained with the \acrshort{ae}} approach and \textit{CNN-1D trained with the  \acrshort{dc}} approach  achieve the performance of (ROC-AUC~=~98.7\%, NMI~=~56.2\%, F1~=~84.8\%), (ROC-AUC~=~99.3\%, NMI~=~58.9\%, F1~=~88.0\%) and (ROC-AUC~=~95.4\%, NMI~=~51.5\%, F1~=~76.0\%), respectively, which shows high quality of cluster assignments in \acrshort{dmm} measured by NMI, and well-shaped clusters reflected by the ROC-AUC and F1 metrics.

The same methods in the \acrshort{dm} setup on the raw I/Q data modality shown in Table \ref{table:oracle_perf} perform significantly worse with (ROC-AUC~=~42.1\%, NMI~=~4.9\%, F1~=~3.4\%), (ROC-AUC~=~50.1\%, NMI~=~2.1\%, F1~=~10.2\%) and (ROC-AUC~=~50.1\%, NMI~=~2.4\%, F1~=~10.1\%) for \textit{PCA, \acrshort{kan} trained by \acrshort{ae}} and \textit{CNN-1D  trained by \acrshort{dc}} methods, respectively. This can be attributed to the prevalent semantic information in samples, so self-supervised approaches tend to separate messages instead of emitters.  

The results in Table~\ref{table:oracle_perf_const} for using the 2D constellation data modality in the \acrshort{dm} scenario for the same methods indicate notable performance improvement across the three metrics: (ROC-AUC~=~60.1\%, NMI~=~34.3\%, F1~=~17.1\%), (ROC-AUC~=~70.6\%, NMI~=~41.1\%, F1~=~38.7\%) and (ROC-AUC~=~65.8\%, NMI~=~43.3\%, F1~=~24.1\%). This confirms that the 2D constellation data modality reduces the presence of semantic information, allowing self-supervised approaches to better separate emitters.
Overall, 2D constellation data modality shows significant improvement in performance under \acrshort{dm} due to the mitigation of different messages difference as described in Sec. \ref{subsec:same_and_diff_scenraious}. 

\subsection{The impact of \acrlong{flm} }

Table~\ref{table:wisig_perf} shows that in the \acrshort{sm} setup, the \acrshort{kan} architecture slightly but consistently outperforms CNN-1D  across all learning paradigms (\acrshort{ae}, \acrshort{dc}, and \acrshort{cl}). 
PCA achieves comparable performance while requiring nearly an order of magnitude fewer parameters and FLOPs, 
highlighting its efficiency as a lightweight baseline. In the \acrshort{dm} scenario using the raw I/Q data modality, Table~\ref{table:oracle_perf}, 
the \acrshort{kan}-based models exhibit weaker cluster assignment quality compared to the CNN-based counterparts, 
whereas PCA underperforms both deep learning approaches across all metrics. For the 2D constellation data modality (Table~\ref{table:oracle_perf_const}), 
\acrshort{kan} initialized with SVDweights and trained via \acrshort{dc} achieves the best performance, substantially surpassing other architectures.
CNN-2D models exhibit stable behavior with SVD initialization when trained using \acrshort{cl}. 
PCA continues to provide a strong baseline with competitive clustering quality while maintaining the lowest computational cost. Comparing CNN-1D and \acrshort{kan} models for the raw I/Q data modality, the latter are several times more demanding in both parameters and FLOPs. However, for the 2D constellation data modality, \acrshort{kan}s are more effective in terms of FLOPs. 

Specifically, in the \acrshort{sm} scenario from Table~\ref{table:wisig_perf}, 
the \acrshort{kan}-based \acrshort{ae} model achieves the best overall performance, reaching (ROC-AUC~=~99.3\%, NMI~=~59\%, F1~=~88\%), compared to the CNN-1D  variant with (ROC-AUC~=~99.2\%, NMI~=~57\%, F1~=~88\%). 
A similar trend is observed for \acrshort{dc} and \acrshort{cl} paradigms, 
where \acrshort{kan} variants consistently outperform their CNN counterparts by on average $2$–$4$ percentage points in ROC-AUC, NMI and F1 metrics. 
In the same table, PCA achieves (ROC-AUC~=~99\%, NMI~=~56\%,F1~=~85\%), 
closely matching deep learning models while using only $10$k parameters and $0.2$M FLOPs, requiring approximately $8$–$30$ times fewer computational resources than CNN or \acrshort{kan} architectures. 

In the \acrshort{dm} scenario from Table~\ref{table:oracle_perf}, the distinction between models diminishes considerably, with all deep learning approaches performing close to the chance level: 
the best ROC-AUC does not exceed $(51.3{\pm}1.7){\%}$ (\acrshort{cl}~CNN-1D ), and F1 values remain around $10{\%}$. 
Here, \acrshort{kan} architectures perform slightly worse in clustering quality (i.e., NMI~=~0.5\% for \acrshort{dc}~\acrshort{kan}) compared to CNN models (i.e., NMI~=~2.4\% for \acrshort{dc}~CNN-1D ). At the same time, PCA drops significantly to (ROC-AUC~=~42\%, NMI~=~5\%, F1~=~3\%), confirming its inability to separate emitters when signal variability is dominated by the message content.

Finally, in the 2D constellation data modality from Table~\ref{table:oracle_perf_const}, 
\acrshort{kan}s initialized with SVD weights and trained by \acrshort{dc} demonstrate a clear advantage, achieving (ROC-AUC~=~71\%, NMI~=~41\%, F1~=~39\%), 
which surpasses all other architectures by at least $4$–$10$ percentage points across metrics. 
In contrast, the best CNN-2D model trained with \acrshort{cl} reaches (ROC-AUC~=~66\%, NMI~=~34\%, F1~=~30\%) 
while PCA achieves (ROC-AUC~=~60\%, NMI~=~34\%, F1~=~17\%), maintaining competitive performance despite minimal complexity. 
Collectively, these results confirm that \acrshort{kan} architectures benefit substantially from SVD-based initialization under \acrshort{dc},
and that PCA remains an efficient and interpretable baseline delivering strong performance relative to its computational footprint.

\begin{figure}[t!]
\centering
\includegraphics[width=0.45\textwidth]{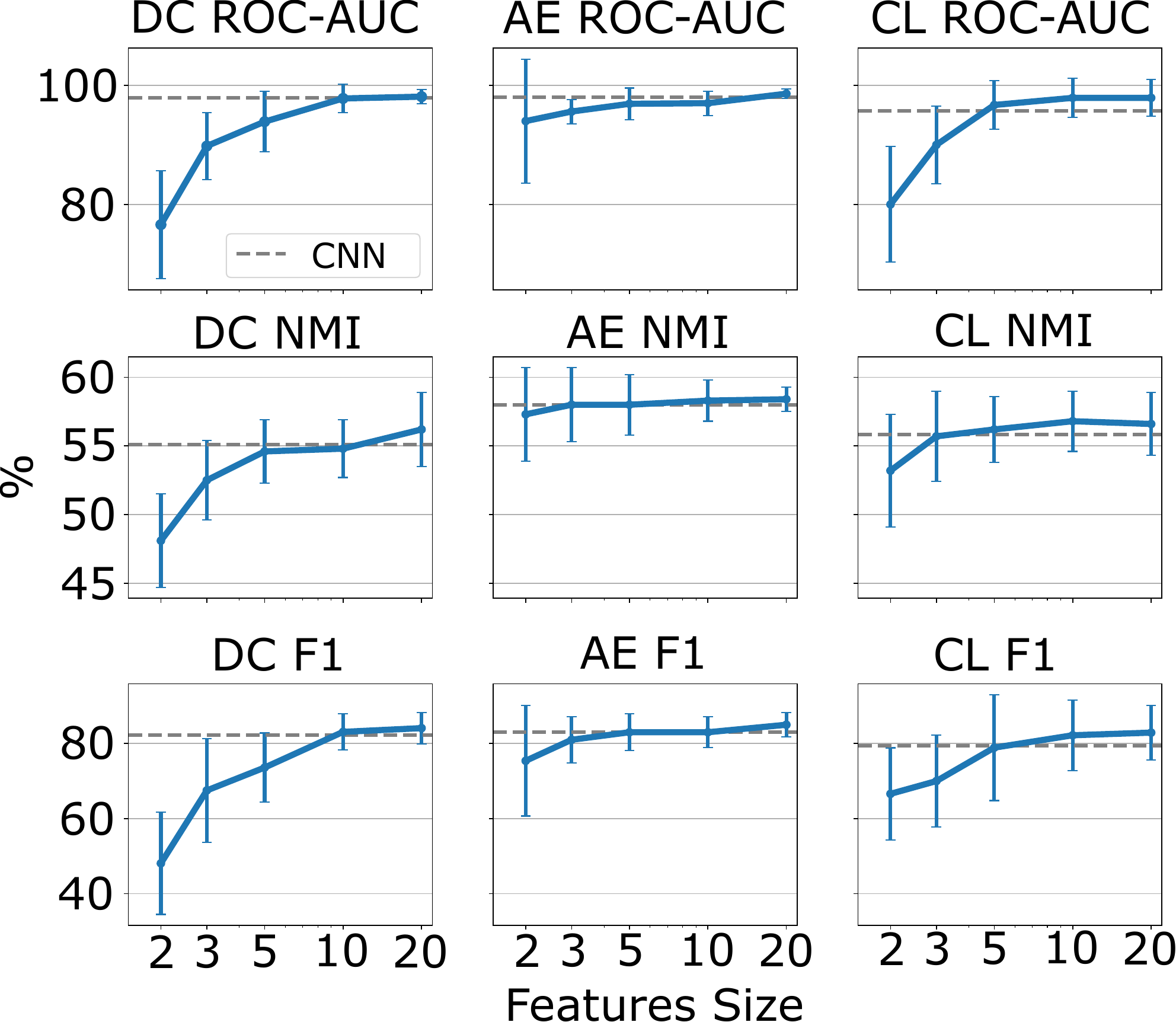} 
\caption{\acrshort{kan}s vs baseline CNN performance on WiSig Dataset. Blue lines represent the average value, blue ticks represent the error bars, and dashed gray lines show the baseline CNN performance.} 
\label{fig:kans_wisig_perf}
\end{figure}

Figure~\ref{fig:kans_wisig_perf} compares \acrshort{kan} and CNN performance on the WiSig dataset across \acrshort{kan} feature sizes and learning paradigms. 
In all cases, \acrshort{kan}s exhibit a clear monotonic improvement as the feature dimensionality increases, most notably under the Deep Clustering (\acrshort{dc}) and Contrastive Learning (\acrshort{cl}) schemes, where ROC-AUC rises from approximately $85\%$ to above $97\%$. NMI improves by nearly $10$ percentage points when moving from $2$ to $20$ features. 
\acrshort{ae}-based \acrshort{kan}s display smaller yet consistent gains, maintaining superior low-dimensional performance compared to other paradigms. 
Across all metrics, \acrshort{kan} models surpass the CNN baseline (dashed gray line) once the feature dimensionality exceeds $5$, confirming that the adaptive spline-based functional representation in \acrshort{kan}s scales more effectively with representational capacity than convolutional filters. 
These results suggest that \acrshort{kan}s can utilize higher-dimensional embeddings to enhance discriminative clustering and improve open-set separability. In contrast, CNNs reach saturation at lower feature sizes due to their limited nonlinearity and fixed receptive-field structure. Summing up, Kolmogorov-Arnold networks (KANs) are interpretable
and have performance comparable to CNNs, a convenient
alternative that suffers from ”black-box” design, across \acrshort{sm}
and \acrshort{dm}. 
\subsection{The impact of the machine learning approach}
In Table~\ref{table:wisig_perf}, it is evident that in the \acrshort{sm} scenario with the WiSig dataset and raw I/Q data modality, the \acrshort{ae} learning approach achieves the best performance, followed by \acrshort{cl}. However, \acrshort{ae}, \acrshort{dc}, and \acrshort{cl} learning approaches evaluated in the \acrshort{dm} setup using the raw I/Q data modality do not show significant differences, as shown in Table~\ref{table:oracle_perf}, with PCA performing the worst. Finally, Table~\ref{table:oracle_perf_const} shows that on the sparse 2D constellation data modality, the \acrshort{dc} approach is the best.

Quantitatively, the results in Table~\ref{table:wisig_perf} confirm the superior performance of the \acrshort{ae} approach in the \acrshort{sm} setup using raw I/Q data. 
The \acrshort{kan}-based \acrshort{ae} achieves the highest average scores across all metrics, with (ROC-AUC~=~99\%, NMI~=~59\%, F1~=~88\%), 
surpassing the best \acrshort{cl} \acrshort{kan} configuration (ROC-AUC~=~98\%, NMI~=~56\%, F1~=~84\%). 
These results indicate that \acrshort{ae} captures the signal structure most effectively when message content remains consistent. 
While PCA provides a strong linear baseline, \acrshort{cl} lags slightly behind, with (ROC-AUC~=~98\%, NMI~=~56\%, F1~=~84\%) for \acrshort{cl} and  (ROC-AUC~=~99\%, NMI~=~56\%, F1~=~85\%) for PCA. 

In contrast, Table~\ref{table:oracle_perf} demonstrates that for the \acrshort{dm} scenario under the raw I/Q data modality, the overall performances of \acrshort{ae}, \acrshort{dc}, and \acrshort{cl} approaches are nearly indistinguishable, 
with ROC-AUC values clustered around $50\%$, NMI between 0.5\% and 10\%, and F1 scores between $9\%$ and $12\%$.

Finally, Table~\ref{table:oracle_perf_const} shows that for the sparse 2D constellation data modality, the \acrshort{dc} approach achieves the best overall results, particularly for the \acrshort{kan} models initialized with SVD weights, 
reaching (ROC-AUC~=~71\%, NMI~=~41\%, F1~=~39\%). 
\acrshort{cl} ranks second with (ROC-AUC~=~66\%, NMI~=~34\%, F1~=~30\%), 
while PCA, although simpler, remains competitive at (ROC-AUC~=~60\%, NMI~=~34\%, F1~=~17\%). Summarizing, for \acrshort{dm}, the best performing configuration is SVD initialized KAN with deep clustering
approach and 2D constellation data modality.

\begin{figure}[t!]
\centering
\includegraphics[width=0.5\textwidth]{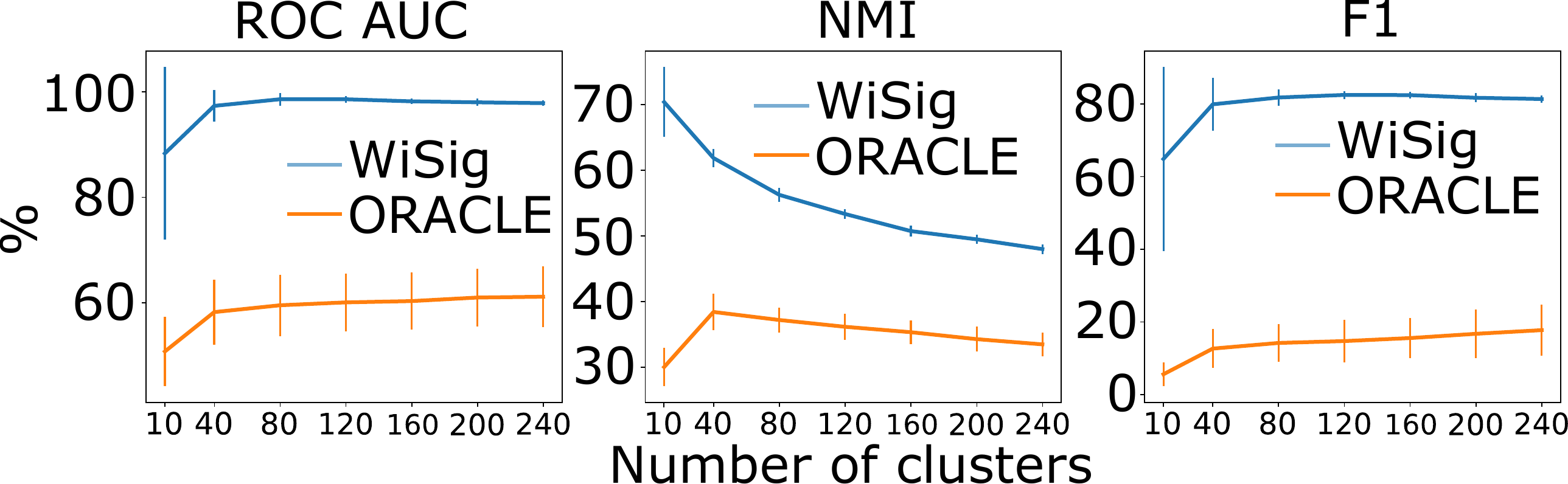}
\caption{Sensitivity of the DMM to the number of clusters. Without loss of generality, we use the PCA baseline with feature size 20.}
\label{fig:clusters}
\end{figure}
\subsection{The impact of the number of clusters in DMM}
Figure~\ref{fig:clusters} shows the DMM's sensitivity to the number of selected clusters using the PCA baseline and measured with ROC-AUC, NMI, and F1 metrics on the ORACLE and WiSig datasets. The results reveal that ROC-AUC and F1 reach a plateau, with only minor changes once the number of clusters becomes sufficiently large, while NMI consistently decreases. The decrease in the NMI metric does not mean the decrease in cluster assignment \cite{mahmoudi2024proof}, because NMI also depends on the number of train emitters. These results indicate that the number of clusters required to reach stability is dataset-dependent. For the WiSig dataset with six emitters, stability is reached at approximately 80 clusters, whereas for the ORACLE dataset with 16 emitters, the stable region begins at approximately 200 clusters.  
\begin{figure*}[t!]
\centering
\hspace*{3cm}
\begin{subfigure}[b]{0.20\textwidth}
    \includegraphics[width=\textwidth]{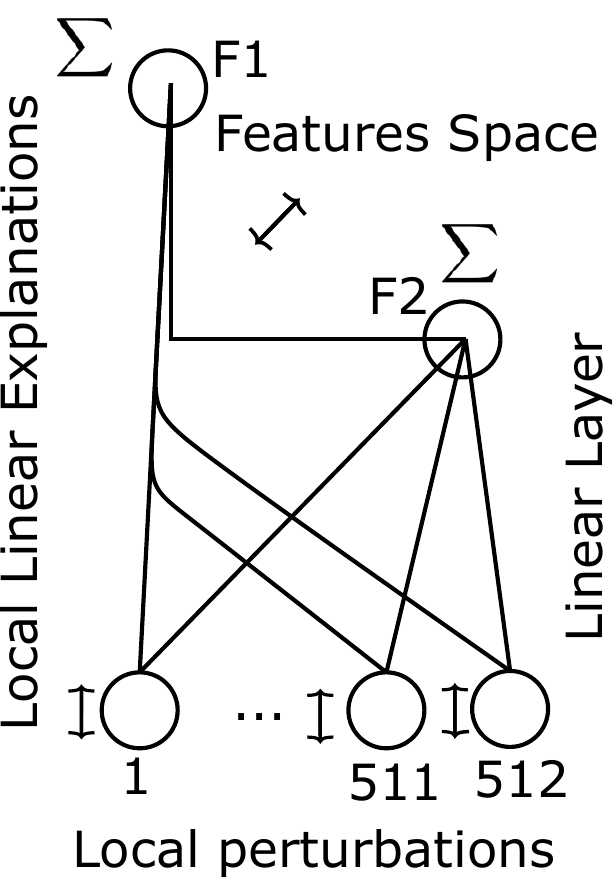}
    \caption{LIME for MLP}
    \label{fig:kans_wisig_inter_lime}
\end{subfigure}
\hfill
\centering
\begin{subfigure}[b]{0.25\textwidth}
\includegraphics[width=\textwidth]{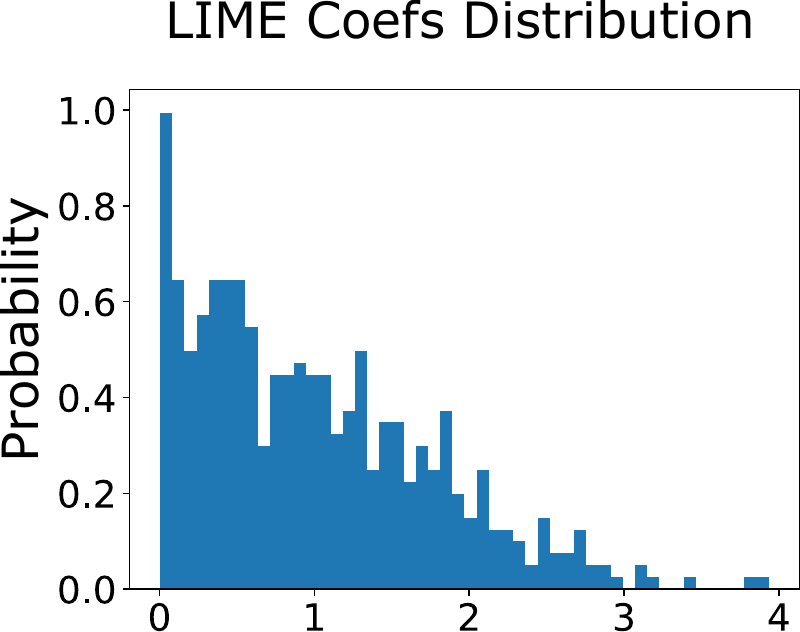}
\caption{LIME coefficients}
\label{fig:kans_wisig_inter_lime_coefs}
\end{subfigure}
\hspace*{3cm}
\\
\centering
\begin{subfigure}[b]{0.25\textwidth}
\includegraphics[width=\textwidth]{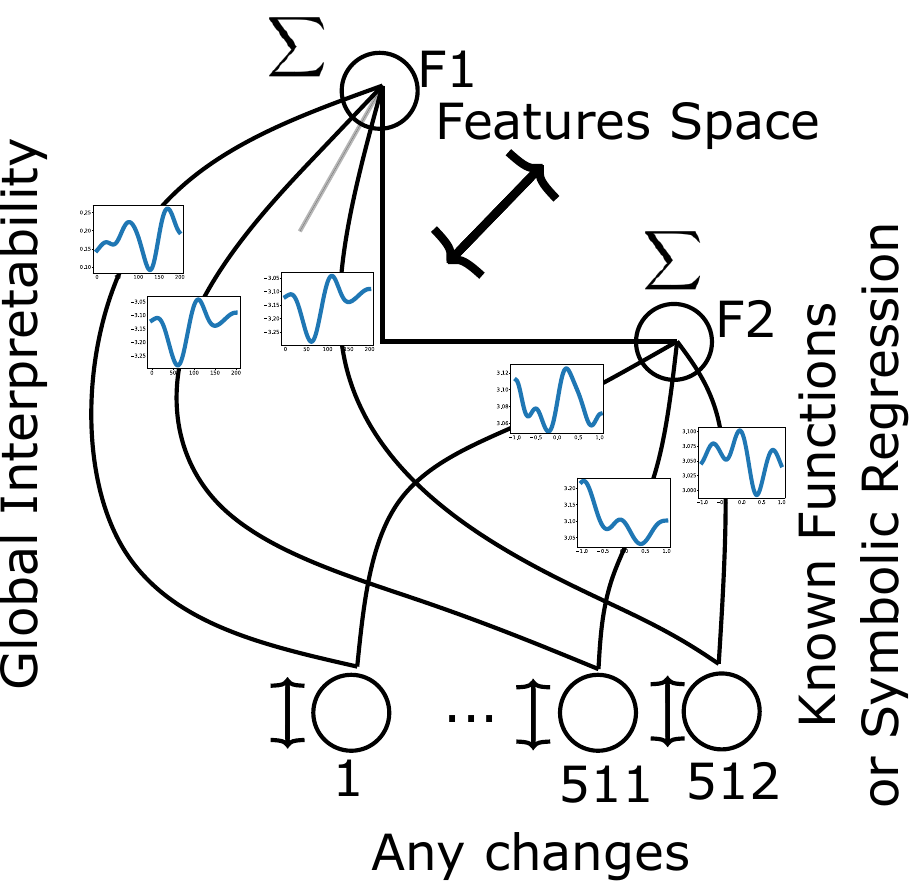}
    \caption{\acrshort{kan} concept}
\label{fig:kans_wisig_inter_scheme}
\end{subfigure}
\hfill
\centering
\begin{subfigure}[b]{0.25\textwidth}
\includegraphics[width=\textwidth]{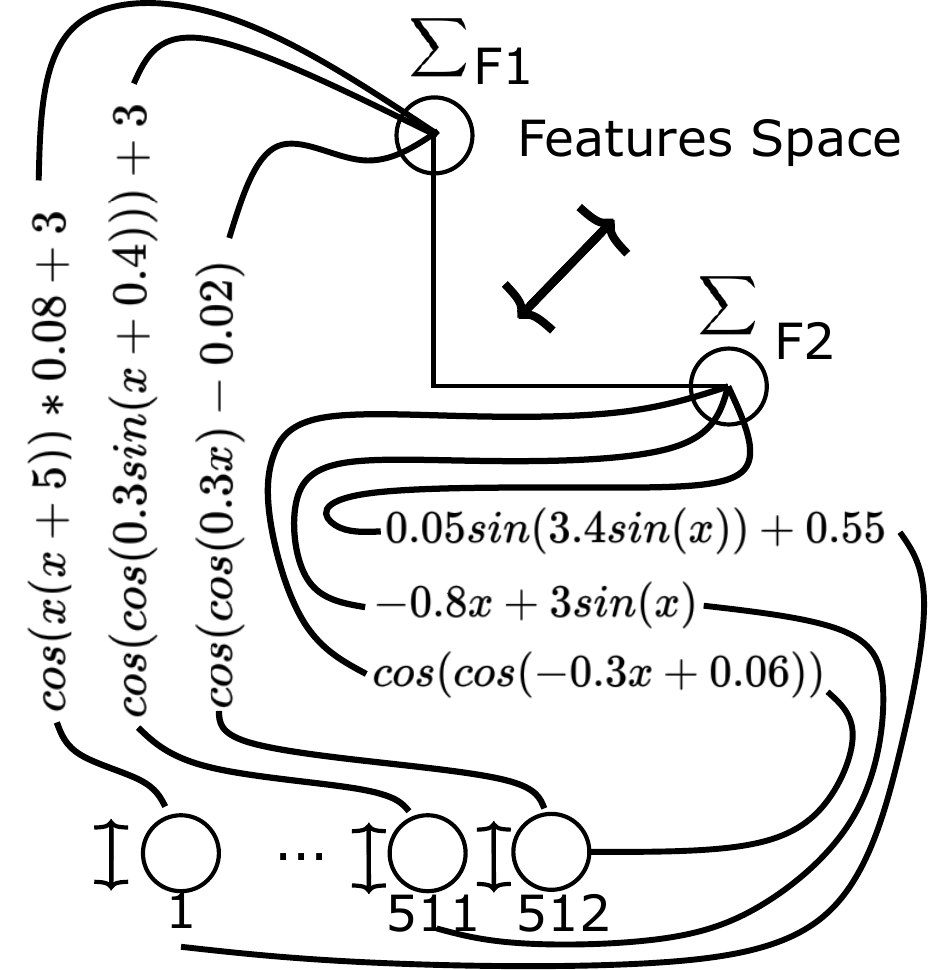}
    \caption{\acrshort{kan} with symbolic regresssion}
    \label{fig:kans_wisig_inter_symb}  
\end{subfigure}
\hfill
\centering
\begin{subfigure}[b]{0.25\textwidth}
\includegraphics[width=\textwidth]{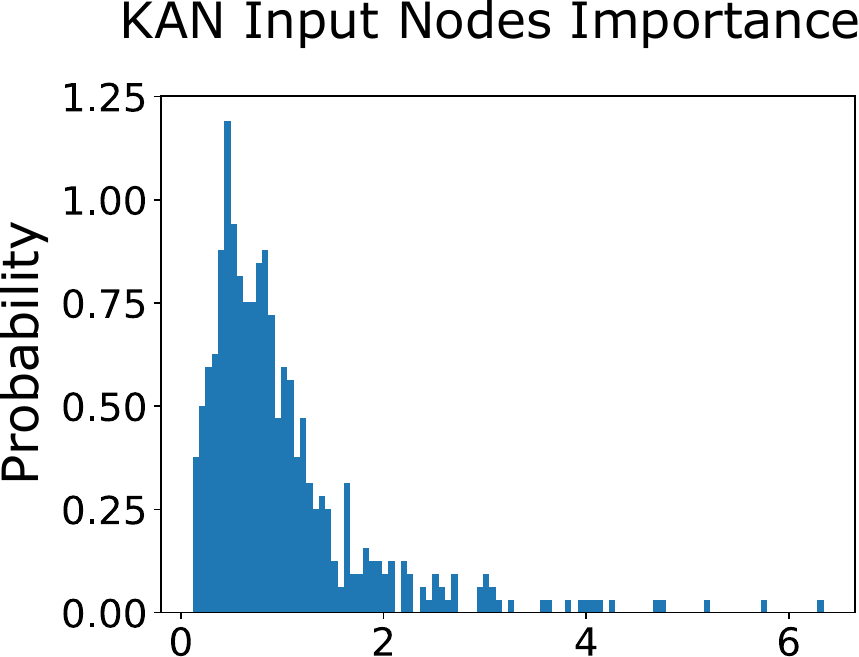}
    \caption{\acrshort{kan} input nodes importance}
\label{fig:kans_wisig_inter_nodes}
\end{subfigure}
\caption{Model interpretability insights for WiSig dataset}
\label{fig:kans_inter_wisig}
\end{figure*}
\subsection{Interpretability evaluation}
Ultimately, we assess the implementation approaches to the proposed concept in terms of interpretability. CNNs incorporate several techniques for interpreting model decisions. One widely used method is LIME, which approximates the model’s behavior around a specific input sample using a linear model:
\begin{equation}
\label{eq:lime}
    M(\delta x + x) \approx \mathbf{W} \delta x + M(x), \quad \text{where } \frac{\|\delta x\|}{\|x\|} \ll 1.
\end{equation}

\noindent Here, $x$ is the input sample, and the matrix $\mathbf{W}$ captures the local linear approximation of the model. LIME can further be used to analyze how specific inputs influence the sample's position in the embedding space, or serve as a locally fully explainable method, especially after pruning non-essential weights from $\mathbf{W}$.

For the WiSig dataset, this concept is illustrated in Figure \ref{fig:kans_wisig_inter_lime}, where a linear layer approximates the CNN’s behavior in the vicinity of a data point. The corresponding distribution of weights from these locally fitted linear models is shown in Figure \ref{fig:kans_wisig_inter_lime_coefs}. However, due to the inherent non-linearity of CNNs, such local interpretability methods cannot generalize across the entire input space. In contrast, Kolmogorov–Arnold Networks provide \textit{global interpretability}. As depicted in Figure \ref{fig:kans_wisig_inter_scheme}, each component of the 2D embedding space is connected to the inputs via known spline functions. The most important connections (black lines) can be approximated using symbolic regression, while less important ones are shown in gray. The possible symbolic regression of spline functions is shown in Figure \ref{fig:kans_wisig_inter_symb}. This structure enables tracing how input perturbations affect internal representations. The distribution of input importance across \acrshort{kan}s is illustrated in Figure \ref{fig:kans_wisig_inter_nodes}, providing a more holistic view of feature influence.

\section{Conclusions}
\label{sec:conc}
This paper presents a comprehensive analysis of the design space for \acrlong{ued}. We explore the key design dimensions with respect to \acrshort{sm} and \acrshort{dm}. Through the lens of the \acrshort{ml} workflow, we investigate the impact of (a) data modality, (b) feature learning module, (c) machine learning approach, and (d) decision-making module. Our theoretical and experimental results reveal several key insights: 1) Under realistic \acrshort{dm} conditions, using a 2D constellation data modality improves  performance by up to 20 p.p.\ in ROC-AUC compared to raw I/Q data.
2) \acrshort{kan} provides an interpretable architecture and performs comparably to CNNs, a “black-box” alternative, across both \acrshort{sm} and \acrshort{dm} settings. 3) For \acrshort{dm}, the best-performing configuration combines an SVD-initialized KAN with a deep clustering approach and 2D constellation data. This setup outperforms a standard KAN (with all other workflow components held constant) by up to 20 p.p.\ in ROC-AUC. 4) In analyzing the decision-making module, we identify the optimal number of clusters required for environments with varying numbers of known emitters. In summary, the results for \acrshort{ued} demonstrate that incorporating \acrshort{kan}-based feature extractors and interpretable representations at both the \acrlong{dmm} and \acrlong{flm} enables the proposed model to generalize effectively to unseen emitters and varying signal conditions, a capability not previously shown in the literature.



\bibliographystyle{IEEEtran}
\bibliography{main}
\end{document}